\documentclass[10pt,journal]{IEEEtran}
\ifCLASSOPTIONcompsoc
  \usepackage[nocompress]{cite}
\else
  \usepackage{cite}
\fi
\ifCLASSINFOpdf
\else
\fi
\usepackage{times}  %Required
\usepackage{helvet}  %Required
\usepackage{courier}  %Required
\usepackage{url}  %Required
\usepackage{graphicx}  %Required
\usepackage{amssymb,amsmath,mathtools, mathrsfs, dsfont}
\usepackage{graphicx, epstopdf}
\usepackage{booktabs}       % professional-quality tables
\usepackage{amsfonts}       % blackboard math symbols
\usepackage{caption}
\usepackage{bm}
\usepackage{color}

\def\kth{$k^{\text{th}}$}
\def\ith{$i^{\text{th}}$}	
\def\jth{$j^{\text{th}}$}

\renewcommand{\vec}[1]{\mathbf{#1}}

\newcommand{\KLD}[2]{D_{\text{KL}}\left( #1\parallel#2\right)}

\def\PoisDist{{\mathscr{P}}}
\def\UniDist{{\mathscr{U}}}
\DeclareMathOperator*{\argmin}{arg\,min}
\DeclareMathOperator*{\argmax}{arg\,max}

\renewcommand{\IEEEbibitemsep}{0pt plus 0.5pt}
\makeatletter
\IEEEtriggercmd{\reset@font\normalfont\fontsize{7.9pt}{8.40pt}\selectfont}
\makeatother
\IEEEtriggeratref{1}

\begin{document}
	\bstctlcite{IEEEexample:BSTcontrol}
\title{Neural Network Detection of Data \\Sequences in Communication Systems}

\author{Nariman~Farsad,~\IEEEmembership{Member,~IEEE,}
        and~Andrea~Goldsmith,~\IEEEmembership{Fellow,~IEEE}% <-this % stops a space
\IEEEcompsocitemizethanks{\IEEEcompsocthanksitem Nariman Farsad and Andrea Goldsmith are with the Department
of Electrical Engineering, Stanford University, Stanford,
CA, 94305. Emails: nfarsad@stanford.edu, andrea@wsl.stanford.edu.}
\IEEEcompsocitemizethanks{\IEEEcompsocthanksitem This work was funded by the NSF Center for Science of Information grant NSF-CCF-0939370, and ONR grant N00014-18-1-2191.}
}
\IEEEtitleabstractindextext{%
\begin{abstract}
We consider detection based on deep learning, and show it is possible to train detectors that perform well without any knowledge of the underlying channel models. Moreover, when the channel model is known, we demonstrate that it is possible to train detectors that do not require channel state information (CSI). In particular, a technique we call a sliding bidirectional recurrent neural network (SBRNN) is proposed for detection where, after training, the detector estimates the data in real-time as the signal stream arrives at the receiver. We evaluate this algorithm, as well as other neural network (NN) architectures, using the Poisson channel model, which is applicable to both optical and molecular communication systems. In addition, we also evaluate the performance of this detection method applied to data sent over a molecular communication platform, where the channel model is difficult to model analytically. We show that SBRNN is computationally efficient, and can perform detection under various channel conditions without knowing the underlying channel model. We also demonstrate that the bit error rate (BER) performance of the proposed SBRNN detector is better than that of a Viterbi detector with imperfect CSI as well as that of other NN detectors that have been previously proposed. Finally, we show that the SBRNN can perform well in rapidly changing channels, where the coherence time is on the order of a single symbol duration.
\end{abstract}

\begin{IEEEkeywords}
Machine learning, deep learning, supervised learning, communication systems, detection, optical communication, free-space optical communication, molecular communication.
\vspace{-0.5cm}
\end{IEEEkeywords}}

\maketitle

\IEEEdisplaynontitleabstractindextext

% \fi
\IEEEpeerreviewmaketitle

%\IEEEraisesectionheading{\section{Introduction}\label{sec:introduction}}
\section{Introduction}\label{sec:introduction}

\IEEEPARstart{O}{ne} of the important modules in reliable recovery of data sent over a communication channel is the detection algorithm, where the transmitted signal is estimated from a noisy and corrupted version observed at the receiver. The design and analysis of this module has traditionally relied on mathematical models that describe the transmission process, signal propagation, receiver noise, and many other components of the system that affect the end-to-end signal transmission and reception. Most communication systems today convey data by embedding it into electromagnetic (EM) signals, which lend themselves to tractable channel models based on a simplification of Maxwell's equations. However, there are cases where tractable mathematical descriptions of the channel are elusive, either because the EM signal propagation is very complicated or when it is  poorly understood. In addition, there are communication systems that do not use EM wave signalling and the corresponding communication channel models may be unknown or mathematically intractable. Some examples of the latter are underwater communication using acoustic signals \cite{sto09} as well as molecular communication, which relies on chemical signals to interconnect tiny devices with sub-millimeter dimensions in environments such as inside the human body \cite{mor06,aky08,eckBook,far16ST}. 

Even when the underlying channel models are known, since the channel conditions may change with time, many model-based detection algorithms rely on the estimation of the instantaneous channel state information (CSI) (i.e., channel model parameters) for detection. Typically, this is achieved by transmitting and receiving a predesigned pilot sequence, which is known by the receiver, for estimating the CSI. However, this estimation process entails overhead that decreases the data transmission rate. Moreover, the accuracy of the estimation may also affect the performance of the detection algorithm. 

\begin{figure*}
	\centering
	\includegraphics[width=0.7\textwidth,keepaspectratio]{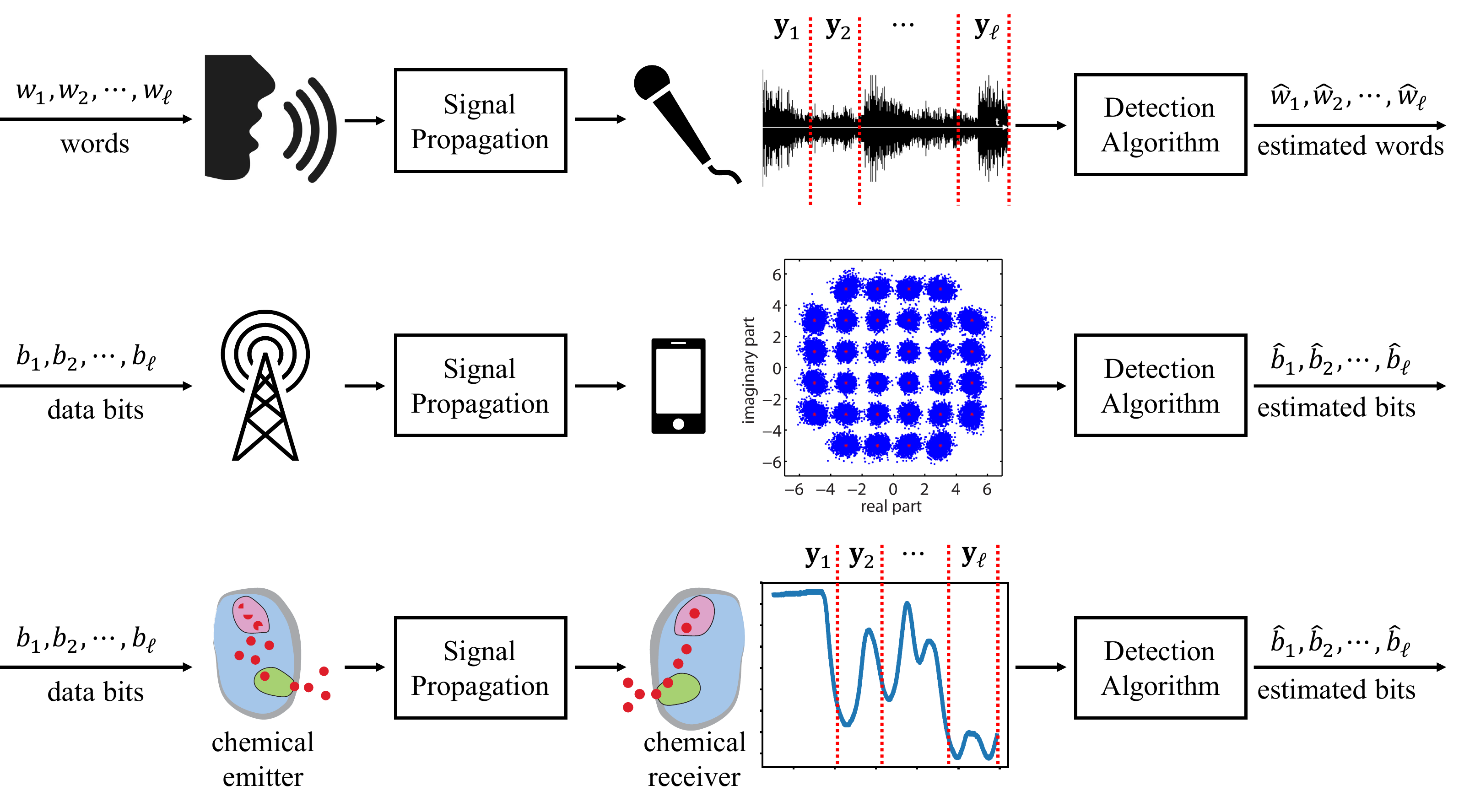}	
	\caption{\label{fig:commSpeechSystem} Similarities between speech recognition and digital communication systems.}
	\vspace{-0.3cm}
\end{figure*}
In this paper, we investigate how different techniques from artificial intelligence and deep learning \cite{lec15,goodfellowBook,ibn00} can be used to design detection algorithms for communication systems that learn directly from data. We show that these algorithms are robust enough to perform detection under changing channel conditions, without knowing the underlying channel models or the CSI. This approach is particularly effective in emerging communication technologies, such as molecular communication, where accurate models may not exist or are difficult to derive analytically. For example, tractable analytical channel models for signal propagation in molecular communication channels with multiple reactive chemicals have been elusive~\cite{far16SPAWC,reactDiffBook,deb11PDEbook}. %Therefore, an approach to design and engineer these systems that does not require analytical channel models is required. 

Some examples of machine learning tools applied to design problems in communication systems include multiuser detection in code-division multiple-access (CDMA) systems \cite{aaz92,mit94,jua06,isi07}, decoding of linear codes \cite{nac16}, design of new modulation and demodulation schemes \cite{dor17,osh16}, detection and channel decoding \cite{nac17b,nac18decodJSTSP, lia18BPCNNJSTSP,cam17polarDecode,dor17deepAirJSTSP,sam17deepMIMO}, and estimating channel model parameters \cite{lee17,osh17CSI}. A recent survey of machine learning techniques applied to communication systems can be found in \cite{osh17introduction}. The approach taken in most of these previous works was to use machine learning to improve one component of the communication system based on the knowledge of the underlying channel models. 

Our approach is different from prior works since we assume that the mathematical models for the communication channel are completely unknown. This is motivated by the recent success in using deep neural networks (NNs) for end-to-end system design in applications such as image classification \cite{kri12imagenet,he16resnet}, speech recognition \cite{hin12, gra14towards,amo16deep}, machine translation \cite{bah14,cho14learning}, and bioinformatics \cite{li16}. For example, Figure~\ref{fig:commSpeechSystem} highlights some of the similarities between speech recognition, where deep NNs have been very successful at improving the detector's performance, and digital communication systems for wireless and molecular channels. As indicated in the figure, for speech processing, the transmitter is the speaker, the transmission symbols are words, and the carrier signal is acoustic waves. At the receiver the goal of the detection algorithm is to recover the sequence of transmitted words from the acoustic signals that are received by the microphone. Similarly, in communication systems, such as wireless or molecular communications, the transmitted symbols are bits and the carrier signals are EM waves or chemical signals. At the receiver the goal of the detection algorithm is to detect the transmitted bits from the received signal. One important difference between communication systems and speech recognition is the size of transmission symbol set, which is significantly larger for speech.  

Motivated by this similarity, in this work we investigate how techniques from deep learning can be used to train a detection algorithm from samples of transmitted and received signals. We demonstrate that, using known NN architectures such as a recurrent neural network (RNN), it is possible to train a detector without any knowledge of the underlying system model. In this approach, the receiver goes through a training phase where a NN detector is trained using known transmission signals. We also propose a real-time NN sequence detector, which we call the {\em sliding bidirectional RNN (SBRNN) detector}, that detects the symbols corresponding to a data stream as they arrive at the destination. We demonstrate that if the SBRNN detector or the other NN detectors considered in this work are trained using a diverse dataset that contains sequences transmitted under different channel conditions, the detectors will be robust to changing channel conditions, eliminating the need for instantaneous CSI estimation for the specific channels considered in this work.

At first glance, the training phase in this approach may seem like an extra overhead. However, if the underlying channel models are known, then the models could be used off-line to generate training data under a diverse set of channel conditions. We demonstrate that using this approach, it is possible to train our SBRNN algorithm such that it would not require any instantaneous CSI. %Note that many detection algorithms in wireless communication must estimate the instantaneous CSI prior to detection. For example, an equalizer typically learns its channel parameters through a training sequence. We also demonstrate that if the underlying channel models are unknown or difficult to obtain, then NN detectors can be trained directly using measured experimental data or field measurements. Finally, 
Another important benefit of using NN detectors in general is that they return likelihoods for each symbol. These likelihoods can be fed directly from the detector into a soft decoding algorithm such as the belief propagation algorithm without requiring a dedicated module to convert the detected symbols into likelihoods.

To evaluate the performance of NN detectors, we first use the Poisson channel model, a common model for optical channels and  molecular communication channels \cite{gha12OpticalBook,gon15,ami15,jam16,jam17ISIT, jam17}. We use this model to compare the performance of the NN detection to the Viterbi detector (VD). We show that for channels with long memories the SBRNN detection algorithm is computationally more efficient than the VD. Moreover, the VD requires CSI estimation, and its performance can degrade if this estimate is not accurate, while the SBRNN detector can perform detection without the CSI, even in a channel with changing conditions. We show that the bit error rate (BER) performance of the proposed SBRNN is better than the VD with CSI estimation error and it outperforms other well-known NN detectors such as the RNN detector. As another performance measure, we use the experimental data collected by the molecular communication platform presented in \cite{far17ExpPlat}. The mathematical models underlying this experimental platform are currently unknown.  We demonstrate that the proposed SBRNN algorithm can be used to train a sequence detector directly from limited measurement data. We also demonstrate that this approach perform significantly better than the detector used in previous experimental demonstrations~\cite{far13,koo16}, as well as other NN detectors. %To demonstrate the practicality of the proposed scheme, we implement the trained NN detectors as part of a text messaging service on our platform. The text message could be of any length, and we are able to reliably transmit and receive messages in real-time at 2 bps. This data rate is an order of magnitude higher than previous systems \cite{far13,koo16}.

\begin{figure*}
	\centering
	\includegraphics[width=1\textwidth,keepaspectratio]{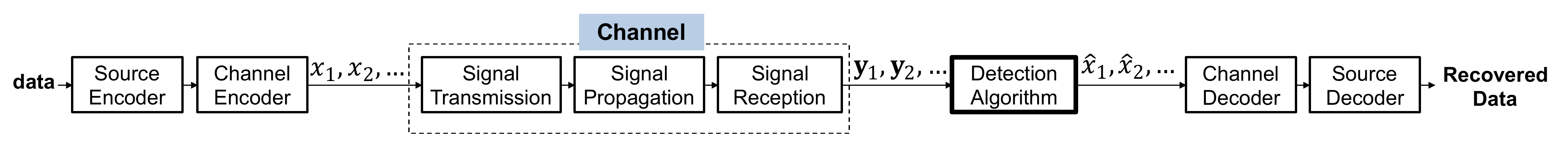}	
	\caption{\label{fig:commSystem} Block diagram for digital communication systems.}
	\vspace{-0.3cm}
\end{figure*}

%Note that although the proposed techniques can be used with any communication system or sequence detection  in speech processing and natural , applying them to molecular communication systems enable many interesting applications. For example, one particular area of interest is in-body communication where bio-sensors, such as synthetic biological devices, constantly monitor the body for different bio-markers for diseases \cite{ata12CM}. Naturally, these biological sensors, which are adapt at detecting biomarkers {\em in vivo} \cite{and06,dan15,slo15}, need to convey their measurements to the outside world. Chemical signaling is a natural solution to this communication problem where the sensor nodes chemically send their measurements to each other or to other devices under/on the skin. The device on the skin is connected to the Internet through wireless technology and can therefore perform complex computations. Thus, the experimental platform we use in this work to validate NN algorithms for signal detection can be used directly to support this important application.   

The rest of the paper is organized as follows. In Section~\ref{sec:probState} we present the problem statement. Then, in Section~\ref{sec:deepDetect}, detection algorithms based on NNs are introduced including the newly proposed SBRNN algorithm. The Poisson channel model and the VD are introduced in Section~\ref{sec:PossinModel}. The performance of the NN detection algorithms are evaluated using this channel model and are compared against the VD in Section~\ref{sec:resultPoiss}. In Section~\ref{sec:resultsExp}, the performance of NN detection algorithms are evaluated using a small data set that is collected via an experimental platform. Concluding remarks are provided in Section~\ref{sec:conclusion}.

\section{Problem Statement}
\label{sec:probState}
In a digital communication system data is converted into a sequence of symbols for transmission over the channel. This process is typically carried out in two steps: in the first step, source coding is used to compress or represent the data using symbols or bits; in the second step, channel coding is used to introduce extra redundant symbols to mitigate the errors that may be introduced as part of the transmission and reception of the data~\cite{vit13DigitalCommBook}. Let $\mathcal{S}=\{s_1, s_2, \cdots, s_m\}$ be the finite set of symbols that could be sent by the transmitter, and $x_k\in\mathcal{S}$ be the \kth~symbol that is transmitted. The channel coding can be designed such that the individual symbols in a long sequence are drawn according to the probability mass function (PMF) $P_X(x)$. %Each symbol is transmitted using a specific signal pattern through a process called modulation, and the released signal then propagates in the environment until it arrives at the receiver. 

The signal that is observed at the destination is noisy and corrupted due to the perturbations introduced as part of transmission, propagation, and reception processes. We refer to these three processes collectively as the {\em communication channel} or simply the {\em channel}. Let the random vector $\vec{y}_k$ of length $\ell$ be the observed signal at the destination during the \kth~transmission. Note that the observed signal $\vec{y}_k$ is typically a vector while the transmitted symbol ${x}_k$ is typically a scalar. A {\em detection algorithm} is then used to estimate the transmitted symbols from the observed signal at the receiver. Let $\hat{x}_k$ be the symbol that is estimated for the \kth~transmitted symbol ${x}_k$. After detection, the estimated symbols are passed to a channel decoder to correct some of the errors in detection, and then to a source decoder to recover the data. All the components of a communication system, shown in Figure~\ref{fig:commSystem}, are designed to ensure reliable data transfer.

Typically, to design these modules, mathematical channel models are required, which describe the relationship between the transmitted symbols and the observed signal through 
\begin{align}
	\label{eq:chanModel}
	P_{\text{model}}(\vec{y}_1, \vec{y}_{2}, \cdots \mid x_1, x_2, \cdots ; \mathbf{\Theta}),
\end{align} 
where $\mathbf{\Theta}$ are the model parameters. Some of these parameters can be static (constants that do not change with channel conditions) and some of them can dynamically change with channel conditions over time. In this work, model parameters are considered to be the parameters that change with time. Hence, we use the terms model parameter and instantaneous CSI interchangeably. Using this model, the detection can be performed through symbol-by-symbol detection, where $\hat{x}_k$ is estimated from $\vec{y}_k$, or using sequence detection where the sequence $\hat{x}_k, \hat{x}_{k-1}, \cdots, \hat{x}_1$ is estimated from the sequence $\vec{y}_k, \vec{y}_{k-1}, \cdots, \vec{y}_1$\footnote{Note that the sequence of symbols $\hat{x}_k, \hat{x}_{k-1}, \cdots, \hat{x}_1$ can also be estimated from $\vec{y}_{k+\ell}, \vec{y}_{k+\ell-1}, \cdots, \vec{y}_1$ for some integer $\ell$. However, to keep the notation simpler, without loss of generality we assume $\ell=0$.}. As an example, for a simple channel with no intersymbol interference (ISI), given by the channel model $P_{\text{model}}(\vec{y}_k \mid x_k;\mathbf{\Theta})$, and a known PMF for the transmission symbols $P_X(x)$, a maximum a posteriori estimation (MAP) algorithm can be devised as
\begin{align}
	\label{eq:egMAP}
	\hat{x}_k = \argmax_{x \in \mathcal{S}} P_{\text{model}}(\vec{y}_k \mid x ; \mathbf{\Theta})P_X(x).
\end{align}
Therefore for detection, both the model and the parameters of the model $\mathbf{\Theta}$, which may change with time, are required. For this reason, many detection algorithms periodically estimate the model parameters (i.e., the CSI) by transmitting known symbols and then using the observed signals at the receiver for CSI estimation \cite{dah13LTEbook}. This extra overhead leads to a decrease in the data rate. One way to avoid CSI estimation is by using blind detectors. These detectors typically assume a particular probability distribution over $\mathbf{\Theta}$, and perform the detection without estimating the instantaneous CSI at the cost of higher probability of error. However, estimating the joint distribution over all model parameters $\mathbf{\Theta}$ can also be difficult, requiring a large amount of measurement data under various channel conditions. One of the problems we consider in this work is whether NN detectors can learn this distribution during training, or learn to simultaneously estimate the CSI and detect the symbols. This approach results in a robust detection algorithm that performs well under different and changing channel conditions without any knowledge of the channel models or their parameters.     

When the underlying channel models do not lend themselves to computationally efficient detection algorithms, or are partly or completely unknown, the best approach to designing detection algorithms is unclear. For example, in communication channels with memory, the complexity of the optimal VD increases exponentially with memory length, and quickly becomes infeasible for systems with long memory. Note that the VD also relies on the knowledge of the channel model in terms of its input-output transition probability.  As another example, tractable channel models for molecular communication channels with multiple reactive chemicals are unknown \cite{far16SPAWC,reactDiffBook, deb11PDEbook}. We propose that in these scenarios, a data driven approach using deep learning is an effective way to train detectors to determine the transmitted symbols directly using known transmission sequences. 

\begin{figure*}[t]
	\centering
	\includegraphics[width=0.9\textwidth,keepaspectratio]{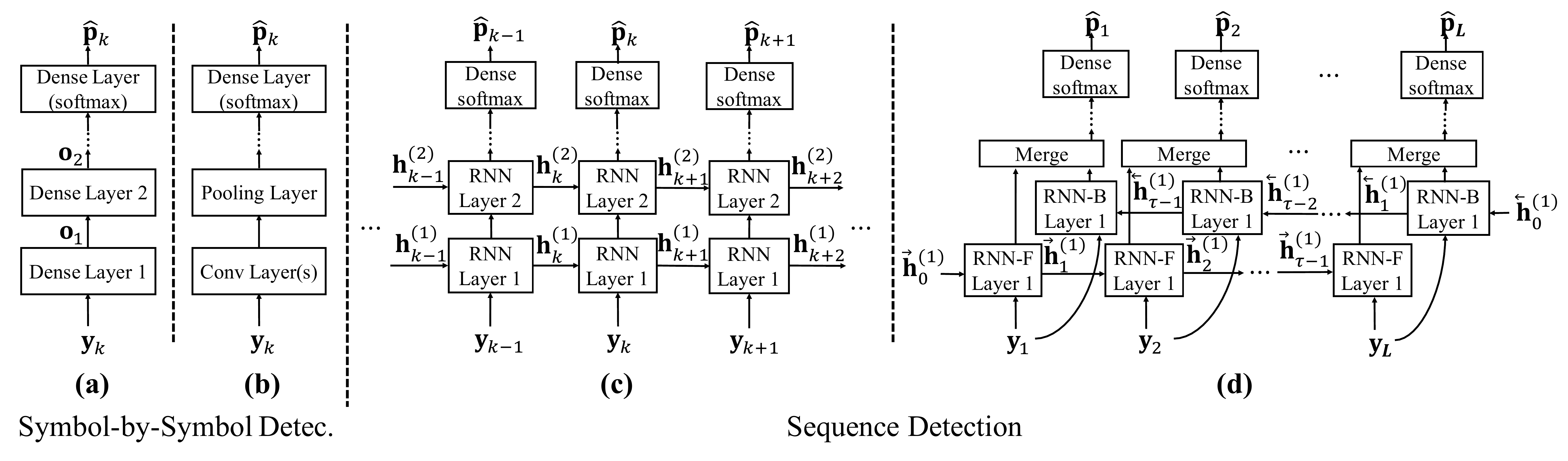}
	\vspace{-0.3cm}
	\caption{\label{fig:NNarchitecture} Different neural network architectures for detection.}
	\vspace{-0.3cm}
\end{figure*}

\section{Detection Using Deep Learning}
\label{sec:deepDetect}
Estimating the transmitted symbol from the received signals can be performed using NN architectures through supervised learning. This is achieved in two phases. First, a training dataset is used to train the NN offline. Once the network is trained, it can be deployed and used for detection. Note that the training phase is performed once offline, and therefore, it is not part of the detection process after deployment. We start this section by describing the training process.

\subsection{Training the Detector}
Let $m=|\mathcal{S}|$ be the cardinality of the symbol set, and let $\vec{p}_k$ be the one-of-$m$ representation of the symbol transmitted during the \kth~transmission, given by
\begin{align}
\label{eq:actualPMF}
\vec{p}_k = 
\begin{bmatrix}
\mathds{1}(x_k=s_1), 
\mathds{1}(x_k=s_2), 
\cdots, 
\mathds{1}(x_k=s_m) 
\end{bmatrix}^\intercal,
\end{align}
where $\mathds{1}(.)$ is the indicator function. Therefore, the element corresponding to the symbol that is transmitted is 1, and all other elements of $\vec{p}_k$ are 0. Note that this is also the PMF of the transmitted symbol during the \kth~transmission where, at the transmitter, with probability 1, one of the $m$ symbols is transmitted. Also note that the length of the vector $\vec{p}_k$ is $m$, which may be different from the length of the vector of the observation signal $\vec{y}_k$ at the destination.

The detection algorithm goes through two phases. In the first phase, known sequences of symbols from $\mathcal{S}$ are transmitted repeatedly and received by the system to create a set of training data. The training data can be generated by selecting the transmitted symbols randomly according to a PMF, and generating the corresponding received signal using mathematical models, simulations, experimental measurements, or field measurements. Let $\vec{P}_K = [\vec{p}_1,\vec{p}_2,\cdots,\vec{p}_K]$ be a sequence of $K$ consecutively transmitted symbols (in the one-of-$m$ encoded representation), and $\vec{Y}_K = [\vec{y}_1,\vec{y}_2,\cdots,\vec{y}_K]$ the corresponding sequence of observed signals at the destination. Then, the training dataset is represented by
\begin{align}
	\{(\vec{P}^{(1)}_{K_1},\vec{Y}^{(1)}_{K_1}),(\vec{P}^{(2)}_{K_2},\vec{Y}^{(2)}_{K_2}), \cdots, (\vec{P}^{(n)}_{K_n},\vec{Y}^{(n)}_{K_n}) \},
\end{align}
which consists of $n$ training samples, where the \ith~sample has $K_i$ consecutive transmissions. 

This dataset is then used to train a deep NN classifier that maps the received signal $\vec{y}_k$ to one of the transmission symbols in $\mathcal{S}$. The input to the NN can be the raw observed signals $\vec{y}_k$, or a set of features $\vec{r}_k$ extracted from the received signals. The NN outputs are the vectors $\hat{\vec{p}}_k=\text{NN}(\vec{y}_k;\mathcal{W})$, where $\mathcal{W}$ are the parameters of the NN. Using the above interpretation of $\vec{p}_k$ as a probability vector, $\hat{\vec{p}}_k$ are the estimations of the probability of $x_k$ given the observations and the parameters of the NN. Note that this output is also useful for soft decision channel decoders (i.e., decoders where the decoder inputs are PMFs), which are typically the next module after detection as shown in Figure~\ref{fig:commSystem}. If channel coding is not used, the symbol is estimated using $\hat{x}_k=\argmax_{x_k\in\mathcal{S}} \hat{\vec{p}}_k$.

During the training, known transmission sequences of symbols are used to find the optimal set of parameters for the NN $\mathcal{W}^*$ such that 
\begin{align}
	\mathcal{W}^* = \argmin_{\mathcal{W}} \mathscr{L}(\vec{p}_k,\hat{\vec{p}}_k),
\end{align}
where $\mathscr{L}$ is the loss function. This optimization algorithm is typically solved using the training data, variants of stochastic gradient decent, and back propagation \cite{goodfellowBook}. Since the output of the NN is a PMF, the cross-entropy loss function can be used for this optimization \cite{goodfellowBook}: 
\begin{align}
\label{eq:lossSymbSymb}
\mathscr{L_{\text{cross}}} = H(\vec{p}_k,\hat{\vec{p}}_k) = H(\vec{p}_k) + \KLD{\vec{p}_k}{\hat{\vec{p}}_k},
\end{align} 
where $H(\vec{p}_k,\hat{\vec{p}}_k)$ is the cross entropy between the correct PMF and the estimated PMF, and $\KLD{.}{.}$ is the Kullback-Leibler divergence \cite{cover-book}. Note that minimizing the loss is equivalent to minimizing the cross-entropy or the Kullback-Leibler divergence distance between the true PMF and the one estimated based on the NN. 
%The categorical cross-entropy loss function can be written as
%%Alternatively, \eqref{eq:lossSymbSymb} and \eqref{eq:lossSeqSeq} can be written as
%%
%\begin{align}
%\mathcal{L_{\text{symb}}} &= -\log\bigg(\hat{\vec{x}}_k [x_k]\bigg),\label{eq:likelihoodSymb}\\
%\mathcal{L_{\text{seq}}} &= -\sum_{k=1}^{\tau} \log\bigg(  \hat{\vec{x}}_k [x_k]\bigg), \label{eq:likelihoodSeq}
%\end{align} 
%%
%where $\hat{\vec{x}}_k [x_k]$ indicates the element of $\hat{\vec{x}}_k$ corresponding to symbol that was actually transmitted, i.e., $x_k$. 
%%Note that minimizing \eqref{eq:likelihoodSymb} and \eqref{eq:likelihoodSeq} 
It is also equivalent to maximizing the log-likelihoods. Therefore, during the training, known transmission data are used to train a detector that {\em maximizes log-likelihoods}.  Using Bayes' theorem, it is easy to show that minimizing the loss is equivalent to maximizing \eqref{eq:egMAP}. 
%Therefore, deep learning can be a powerful tool for designing detection algorithms for communication systems, especially when the underlying channel models are unknown. 
We now discuss how several well-known NN architectures can be used for symbol-by-symbol detection and for sequence detection.

\subsection{Symbol-by-Symbol Detectors}
%\subsubsection{Symbol-by-Symbol Detection}
The most basic NN architecture that can be employed for detection uses several fully connected NN layers followed by a final softmax layer \cite{lec15,goodfellowBook}. 
%In this network, the \ith~layer is given by
%\begin{align}
%\label{eq:denseLayer}
%\vec{o}_i = f(\vec{W}_i \vec{o}_{i-1} + \vec{b}_i),
%\end{align}
%where $f()$ is the activation function, $\vec{W}_i$ and $\vec{b}_i$ are the weight and bias parameters, and $\vec{o}_i$ is the output of the \ith~layer. 
The input to the first layer is the observed signal $\vec{y}_k$ or the feature vector $\vec{r}_k$, which is selectively extracted from the observed signal through preprocessing. The output of the final layer is of length $m$ (i.e., the cardinality the symbol set), and the activation function for the final layer is the softmax activation. 
%given by
%\begin{align}
%\label{eq:softmax}
%\phi_i(\vec{z})=\frac{e^{z_i}}{\sum_j e^{z_j}}.
%\end{align} 
This ensures that the output of the layer $\hat{\vec{p}}_k$ is a PMF. Figure \ref{fig:NNarchitecture}(a) shows the structure of this NN. 

A more sophisticated class of NNs that is used in processing complex signals such as images is a convolution neural network (CNN) \cite{law97,kri12,lec15}. Essentially, the CNN is a set of filters that are trained to extract the most relevant features for detection from the received signal. The final layer in the CNN detector is a dense layer with output of length $m$, and a softmax activation function. This results in an estimate $\hat{\vec{p}}_k$ from the set of features that are extracted by the convolutional layers in the CNN. Figure \ref{fig:NNarchitecture}(b) shows the structure of this NN.    

For symbol-by-symbol detection the estimated PMF $\hat{\vec{p}}_k$ is given by
\begin{align}
\footnotesize
\label{eq:estSymbPMF}
\hat{\vec{x}}_k = 
\begin{bmatrix}
P_{\text{NN}}(x_k=s_1|\vec{y}_k),
P_{\text{NN}}(x_k=s_2|\vec{y}_k), 
\cdots,
P_{\text{NN}}(x_k=s_m|\vec{y}_k)
\end{bmatrix}^\intercal,
\end{align}
where $P_{\text{NN}}$ is the probability of estimating each symbol based on the NN model used. 
The better the structure of the NN at capturing the physical channel characteristics based on $P_{\text{model}}$ in \eqref{eq:chanModel}, the better this estimate and the results. %Therefore, it is important to use insights from the physical channel characteristics when designing the network architecture.   

\subsection{Sequence Detectors}
The symbol-by-symbol detector cannot take into account the effects of ISI between symbols\footnote{It is possible to use the received signal from multiple symbols as input to a CNN for detection in the presence of ISI.}. In this case, sequence detection can be performed using recurrent neural networks (RNN) \cite{lec15,goodfellowBook}, which are well established for sequence estimation in different problems such as  neural machine translation \cite{bah14}, speech recognition \cite{hin12}, or bioinformatics \cite{li16}.
The estimated $\hat{\vec{p}}_k$ in this case is given by
\begin{align}
\label{eq:estForwSeqPMF}
\hat{\vec{p}}_k = 
\begin{bmatrix}
P_{\text{RNN}}(x_k=s_1|\vec{y}_k,\vec{y}_{k-1},\cdots,\vec{y}_1)   \\
P_{\text{RNN}}(x_k=s_2|\vec{y}_k,\vec{y}_{k-1},\cdots,\vec{y}_1) \\
\vdots \\
P_{\text{RNN}}(x_k=s_m|\vec{y}_k,\vec{y}_{k-1},\cdots,\vec{y}_1) 
\end{bmatrix},
\end{align}    
where $P_{\text{RNN}}$ is the probability of estimating each symbol based on the NN model used. In this work, we use long short-term memory (LSTM) networks \cite{hoc97}, which have been extensively used in many applications.%Note that the decoder considers the information from previously received signals, encoded in $\vec{a}^{(l)}_{k-1}$ and $\vec{c}^{(l)}_{k-1}$,  as well as the received signal from current symbol for detection.

%which is composed of the following
%\begin{align}
%\vec{i}_k &= \sigma(\vec{W}_{y,i}\vec{y}_k+\vec{W}_{a,i}\vec{a}_{k-1}+\vec{W}_{c,i}\vec{c}_{k-1}+\vec{b}_i),\label{eq:LSTMnode1}\\
%\vec{f}_k &= \sigma(\vec{W}_{y,f}\vec{y}_k+\vec{W}_{a,f}\vec{a}_{k-1}+\vec{W}_{c,f}\vec{c}_{k-1}+\vec{b}_f),\label{eq:LSTMnode2}\\
%\vec{c}_k &= \vec{f}_k \odot \vec{c}_{k-1} + \vec{i}_k  \odot \tanh (\vec{W}_{y,c}\vec{y}_k+\vec{W}_{a,c}\vec{a}_{k-1}+\vec{b}_c), \label{eq:LSTMnode3}\\
%\vec{u}_k &= \sigma(\vec{W}_{y,u}\vec{y}_k+\vec{W}_{a,u}\vec{a}_{k-1}+\vec{W}_{c,u}\vec{c}_{k}+\vec{b}_u), \label{eq:LSTMnode4}\\
%\vec{a}_k &= \vec{u}_k \odot \tanh(\vec{c}_k), \label{eq:LSTMnode5}
%\end{align}
%where $ \odot$ is the element-wise vector multiplication, $\vec{W}$ and $\vec{b}$ are weights and biases, and $\sigma$ is the logistic sigmoid function \cite{goodfellowBook}. 
%The state of each LSTM cell is passed to the next cell in the next time instance. The RNN may have multiple layers, where the output of the LSTM may be passed to other LSTM layers. 

Figure \ref{fig:NNarchitecture}(c) shows the RNN structure. One of the main benefits of this detector is that after training, similar to a symbol-by-symbol detector, it can perform detection on any data stream as it arrives at the receiver. This is because the observations from previous symbols are summarized as the state of the RNN, which is represented by the vector $\vec{h}_k$. Note that the observed signal during the \jth~transmission slot, $\vec{y}_j$ where $j>k$, may carry information about the \kth~symbol $x_k$ due to delays in signal arrival which results in ISI. However, since RNNs are feed-forward only, during the estimation of $\hat{\vec{p}}_k$, the observation signal $\vec{y}_j$ is not considered.  
%Let $\vec{a}^{(l)}_k$ be the output of the final LSTM layer. Then a dense layer with the softmax activation in \eqref{eq:softmax} is used to obtain $\hat{\vec{x}}_k$ as
%\begin{align}
%\hat{\vec{x}}_k = \phi(\vec{W}_{a} \vec{a}^{(l)}_k + \vec{b}_x).
%\end{align}

One way to overcome this limitation is by using bidirectional RNNs (BRNNs), where the sequence of received signals are once fed in the forward direction into one RNN cell and once fed in backwards into another RNN cell \cite{sch97}. The two outputs are then concatenated and may be passed to more bidirectional layers. Figure \ref{fig:NNarchitecture}(d) shows the BRNN structure. For a sequence of length $L$, the estimated $\hat{\vec{p}}_k$ for BRNN is given by
\begin{align}
\label{eq:estBiDirSeqPMF}
\hat{\vec{p}}_k = 
\begin{bmatrix}
\mspace{-3mu} P_{\text{BRNN}}(x_k=s_1|\vec{y}_L,\mspace{-3mu}\vec{y}_{L-1},\mspace{-3mu}\cdots,\mspace{-3mu}\vec{y}_1) \mspace{-3mu} \\
\mspace{-3mu} P_{\text{BRNN}}(x_k=s_2|\vec{y}_L,\mspace{-3mu}\vec{y}_{L-1},\mspace{-3mu}\cdots,\mspace{-3mu}\vec{y}_1) \mspace{-3mu}\\
\vdots \\
\mspace{-3mu} P_{\text{BRNN}}(x_k=s_m|\vec{y}_L,\mspace{-3mu}\vec{y}_{L-1},\mspace{-3mu}\cdots,\mspace{-3mu}\vec{y}_1) \mspace{-3mu}
\end{bmatrix},
\end{align}
where $k \leq L$. In this work we use the bidirectional LSTM (BLSTM) networks \cite{gra05}.

The BRNN architecture ensures that in the estimation of a symbol, future signal observations are taken into account, thereby overcoming the limitations of RNNs. The main trade-off is that as signals from a data stream arrive at the destination, the block length $L$ increases, and the whole block needs to be re-estimated again for each new data symbol that is received. Therefore, this quickly becomes infeasible for long data streams as the length of the data stream can be on the order of tens of thousands to millions of symbols. In the next section we present a new technique to solve this issue.

\subsection{Sliding BRNN Detector}

\begin{figure}
	\centering
	\includegraphics[width=0.8\columnwidth,keepaspectratio]{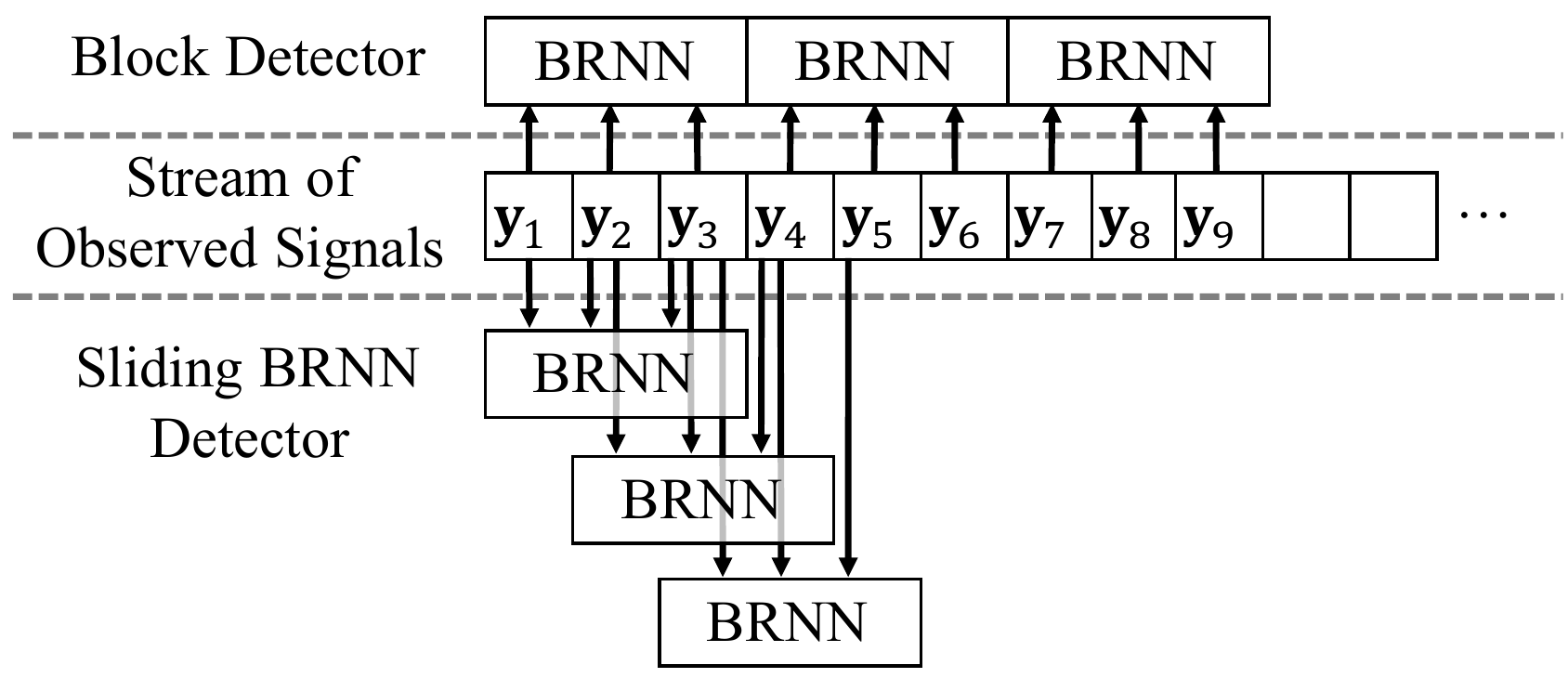}
	\caption{\label{fig:slidingDetector} The sliding BRNN detector.}
	\vspace{-0.3cm}
\end{figure}

Since the data stream that arrives at the receiver can have any arbitrary length, it is not desirable to detect the whole sequence for each new symbol that arrives, as the sequence length could grow arbitrarily large. Therefore, we fix the maximum length of the BRNN. Ideally, the length must be at least the same size as the memory length of the channel. However, if this is not known in advance, the BRNN length can be treated as a hyperparameter to be tuned during training. Let $L$ be the maximum length of the BRNN. Then during training, blocks of $\ell \leq L$ consecutive transmissions are used for training. Note that sequences of different lengths could be used during training as long as all sequence lengths are smaller than or equal to $L$.  After training, the simplest scheme would be to detect the stream of incoming data in fixed blocks of length $\ell \leq L$ as shown in the top portion of Figure~\ref{fig:slidingDetector}. The main drawback here is that the symbols at the end of each block may affect the symbols in the next block, and this relation is not captured in this scheme. Another issue is that $\ell$ consecutive symbols must be received before detection can be performed. The top portion of Figure \ref{fig:slidingDetector} shows this scheme for $\ell=3$.

To overcome these limitations, inspired by some of the techniques used in speech recognition \cite{gra06}, we propose a dynamic programing scheme we call the {\em sliding BRNN (SBRNN) detector}. In this scheme the first $\ell \leq L$ symbols are detected using the BRNN. Then as each new symbol arrives at the destination, the position of the BRNN slides ahead by one symbol. Let the set $\mathcal{J}_k = \{j \mid j \leq k ~\wedge~ j + L > k \}$ be the set of all valid starting positions for a BRNN detector of length $L$, such that the detector overlaps with the \kth~symbol. For example, if $L=3$ and $k=4$, then $j=1$ is not in the set $\mathcal{J}_k$ since the BRNN detector overlaps with symbol positions 1, 2, and 3, and not the symbol position 4. Let $\hat{\vec{p}}^{(j)}_k$ be the estimated PMF for the \kth~symbol, when the start of the sliding BRNN is on $j\in\mathcal{J}_k$. The final PMF corresponding to the \kth~symbol is given by the weighted sum of the estimated PMFs for each of the relevant windows:
\begin{align}
	\label{eq:slidingPMF}
	\hat{\vec{p}}_k = \frac{1}{|\mathcal{J}_k|} \sum_{j\in \mathcal{J}_k} \hat{\vec{p}}^{(j)}_k.
\end{align}
One of the main benefits of this approach is that, after the first $L$ symbols are received and detected, as the signal corresponding to a new symbol arrives at the destination, the detector immediately estimates that symbol. The detector also updates its estimate for the previous $L-1$ symbols dynamically. Therefore, this algorithm is similar to a dynamic programming algorithm.

The bottom portion of Figure~\ref{fig:slidingDetector} illustrates the sliding BRNN detector. In this example, after the first 3 symbols arrive, the PMF for the first three symbols, $i\in\{ 1,2,3\}$, is given by $\hat{\vec{p}}_i = \hat{\vec{p}}^{(1)}_i$. When the 4th symbol arrives, the estimate of the first symbol is unchanged, but for $i\in\{2,3\}$, the second and third symbol estimates are updated as $\hat{\vec{x}}_i = \tfrac{1}{2}(\hat{\vec{x}}^{(1)}_i+\hat{\vec{x}}^{(2)}_i)$, and the 4th symbol is estimated by $\hat{\vec{p}}_4 = \hat{\vec{p}}^{(2)}_4$. Note that although in this paper we assume that the weights of all $\hat{\vec{p}}^{(j)}_k$ are the same (i.e., $\tfrac{1}{|\mathcal{J}_k|}$), the algorithm can use different weights. Moreover, the complexity of the SBRNN increases linearly with the length of the BRNN window, and hence with the memory length.

To evaluate the performance of all these NN detectors, we use both the Poisson channel model (a common model for optical and molecular communication systems) as well as an experimental platform for molecular communication where the underlying model is unknown \cite{far17ExpPlat}. The sequel discusses more details of the Poisson model and experimental platform, and how they were used for performance analysis of our proposed techniques.  %Over the next two section, we describe the Poisson channel models and the experimental platform. %In the next section we describe this platform.

\section{The Poisson Channel Model}
\label{sec:PossinModel}
The Poisson channel has been used extensively to model different communication systems in optical and molecular communication \cite{gha12OpticalBook,gon15,ami15,jam16,jam17ISIT, jam17}. In these systems, information is encoded in the intensity of the photons or particles released by the transmitter and decoded from the intensity of photons or particles observed at the receiver. In the rest of this section, we refer to the photons, molecules, or particles simply as {\em particles}. We now describe this channel, and a VD for the channel.

In our model it is assumed that the transmitter uses on-off-keying (OOK) modulation, where the transmission symbol set is  $\mathcal{S}=\{0,1\}$, and the transmitter either transmits a pulse with a fixed intensity to represent the 1-bit or no pulse to represent the 0-bit. Note that OOK modulation has been considered in many previous works on optical and molecular communication and has been shown to be the optimal input distribution for a large class of Poisson channels \cite{sha90, cao14, far17ISIT}. Later in Section~\ref{subsec:sybolset}, we extend the results to larger symbol sets by considering the general $m$ level pulse amplitude modulation ($m$-PAM), where information is encoded in $m$ amplitudes of the pulse transmissions. Note that OOK is a special case of this modulation scheme with $m=2$.

\begin{figure}
	\centering
	\includegraphics[width=1\columnwidth,keepaspectratio]{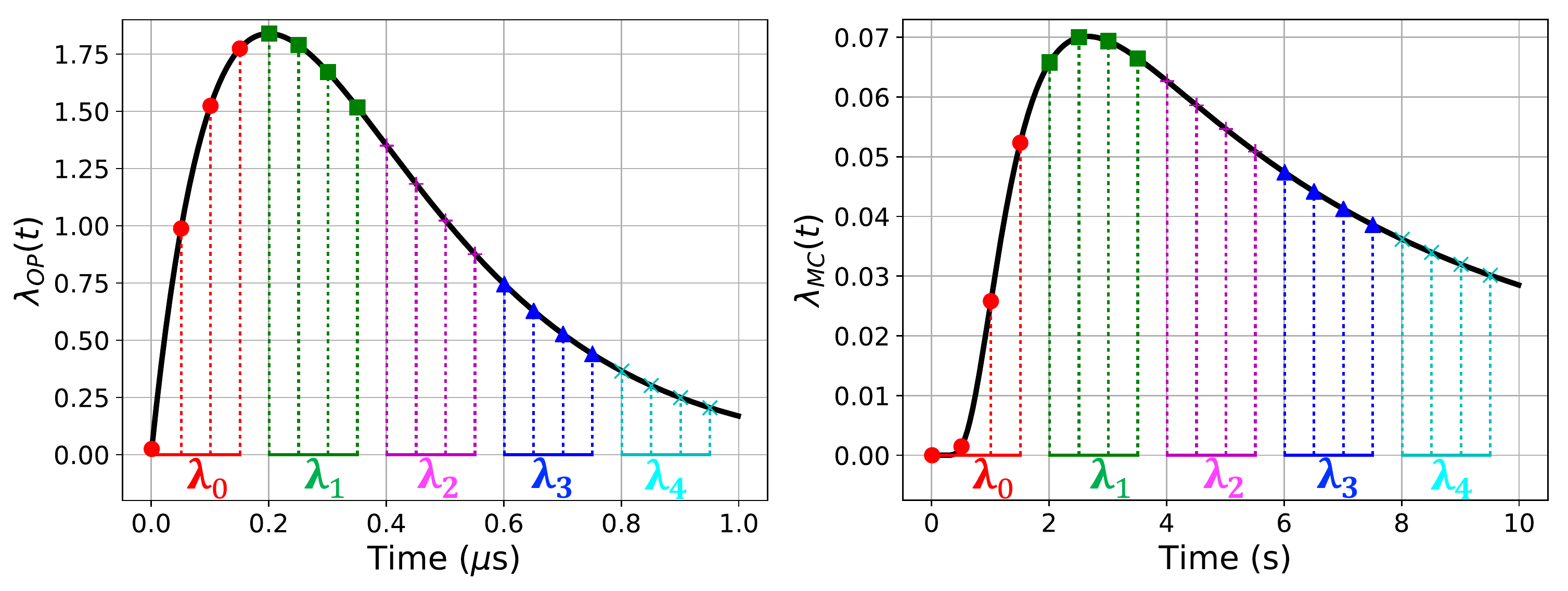}
	\caption{\label{fig:impulseRes} A sample system response for optical and molecular channels. Left: Optical channel with $\lambda(t)$ for $N=1$, $\kappa_{\text{OP}}=1$, $\alpha=2$, $\beta=0.2$, $\tau=0.2$ $\mu$s, and $\omega=20$ MS/s. At $\tau=0.2$ $\mu$s, much of the intensity from the current transmission will arrive during future symbol intervals. Right: Molecular channel with $\kappa_{\text{MO}}=1$, $c=8$, $\mu=40$, $\tau=2$ s, and $\omega=2$ S/s. Molecular channel response has a loner tail than optical channel.}
\end{figure}

Let $\tau$ be the symbol interval, and $x_k\in\mathcal{S}$ the symbol corresponding to the \kth~transmission. We assume that the transmitter can measure the number of particles that arrive at a sampling rate of $\omega$ samples per second. Then the number of samples in a given symbol duration is given by $a = \omega\tau$, where we assume that $a$ is an integer. Let $\lambda(t)$ be the system response to a transmission of the pulse corresponding to the 1-bit. For optical channels, the system response is proportional to the Gamma distribution, and given by \cite{hay07OptiGamma, maj07OptiGamma, din09OptiGamma}:
\begin{align}
\label{eq:impulseResOpti}
\lambda_{\text{OP}}(t) =
\begin{cases}
\kappa_{\text{OP}} \frac{\beta^{-\alpha}t^{\alpha-1}}{\Gamma(\alpha)} \exp(-t/\beta) & t > 0 \\
0 & t\leq 0 
\end{cases},
\end{align} 
where $\kappa_{\text{OP}}$ is the proportionality constant, and $\alpha$ and $\beta$ are parameters of the channel, which can change over time.
For molecular channels, the system response is proportional to the inverse Gaussian distribution \cite{sri12,noe14OptRcv,jam16,jam17ISIT} given by:
\begin{align}
	\label{eq:impulseRes}
	\lambda_{\text{MO}}(t) =
	\begin{cases}
		\kappa_{\text{MO}} \sqrt{\frac{c}{2\pi t^3}} \exp\left[-\frac{c(t-\mu)^2}{2\mu^2t}\right] & t > 0 \\
		0 & t\leq 0 
	\end{cases}, 
\end{align}   
where $\kappa_{\text{MO}}$ is the proportionality constant, and $c$ and $\mu$ are parameters of the channel, which can change over time. 

Since the receiver samples the data at a rate of $\omega$, for $k \in \mathbb{N}$ and $j\in\{1,2,\cdots,a\}$, let 
\begin{align}
	\label{eq:lambdaDefinition}
	\bm{\lambda}_k[j] \triangleq \lambda\left(\frac{j+ka}{\omega}\right) 
\end{align}
be the average intensity observed during the \jth~sample of the \kth~symbol in response to the transmission pulse corresponding to the 1-bit. Figure~\ref{fig:impulseRes} shows the system response for both optical and molecular channels. Although for optical channels the symbol duration is many orders of magnitude smaller than for molecular channels, the system responses are very similar in shape. Some notable differences are a faster rise time for the optical channel, and a longer tail for the molecular channel.  %Note that for optical channels, instead of the inverse Gaussian, $\lambda(t)$ is proportional to the Gamma distribution, which has a shape similar to the inverse Gaussian \cite{din09OptiModel,che10OpticExp}. Therefore, all analysis also holds for the optical channel.

\begin{figure}
	\centering
	\includegraphics[width=0.83\columnwidth,keepaspectratio]{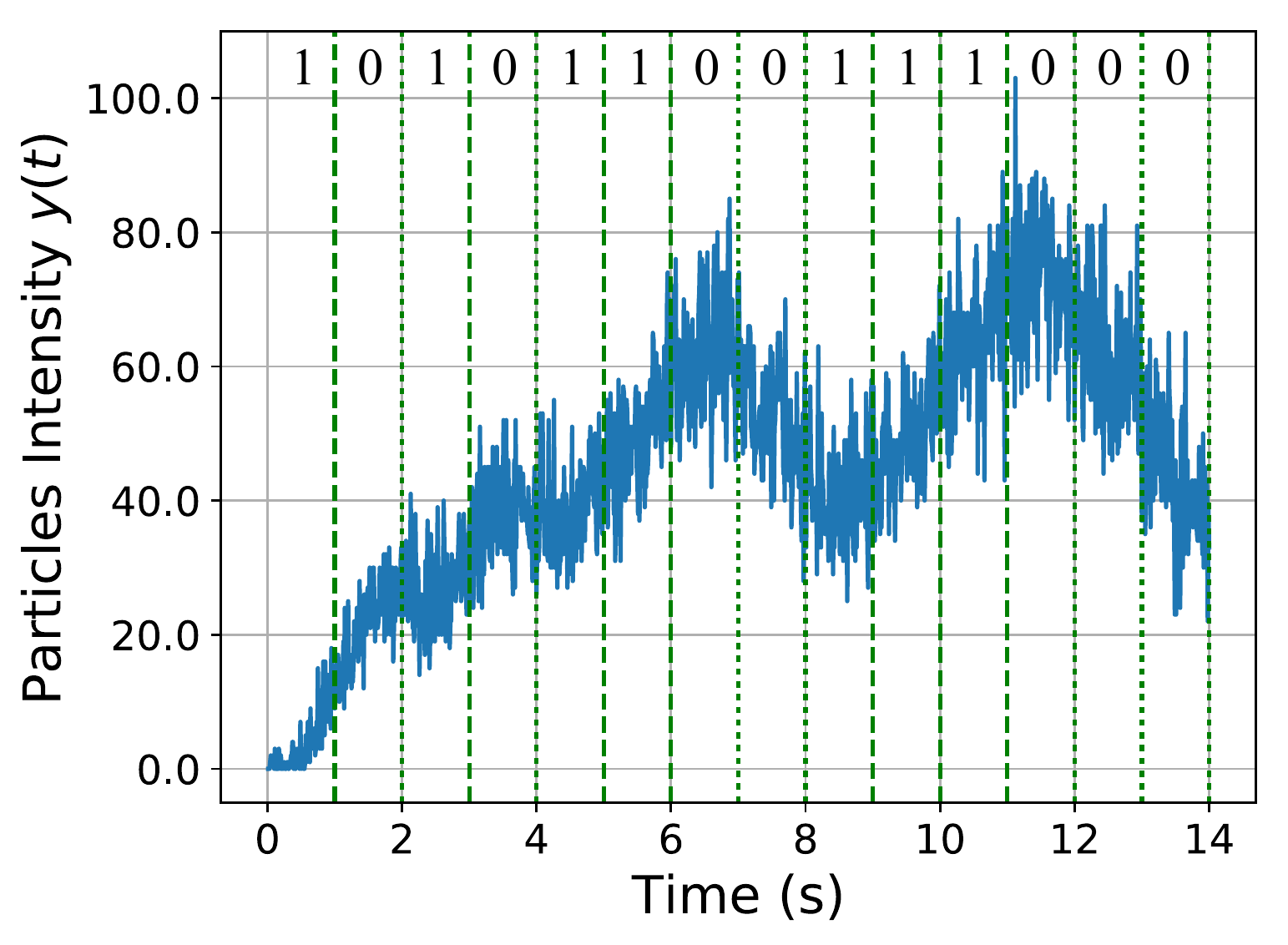}
	\caption{\label{fig:rcvSig} The observed signal for the transmission of the bit sequence 10101100111000 for $\kappa_{\text{MO}}=100$, $c=8$, $\mu=40$, $\tau=1$, $\omega=100$ Hz, and $\eta=1$.}
\end{figure}
The system responses are used to formulate the Poisson channel model. In particular, the intensity that is observed during the \jth~sample of the \kth~symbol is distributed according to
\begin{align}
\label{eq:PoissonMarginal}
\vec{y}_k[j] \sim \PoisDist\left( \sum_{i=0}^{k} x_{k-i} \bm{\lambda}_{i}[j] + \eta \right), 
\end{align}
where $\PoisDist(\xi)=\frac{\xi^y e^{-\xi}}{y!}$ is the Poisson distribution, and $\eta$ is the mean of an independent additive Poisson noise due to background interference and/or the receiver noise\footnote{Note that $\eta$  is the noise term that is typically used in the Poisson channel model. In the optical communication literature this noise is also known as the dark current  \cite{gha12OpticalBook,gon15,ami15}. The noise is due to imperfect receiver, or background noise (due to ambient optical noise or molecules that may exist in the environment).}.  Using this model, the signal that is observed by the receiver, for any sequence of bit transmissions, can be generated as illustrated in Figure~\ref{fig:rcvSig}. This signal has a similar structure to the signal observed using the experimental platform in \cite[see Figure 13]{far13}, although this analytically-modeled signal exhibits more noise.

The model parameters (i.e., the CSI) for the Poisson channel model are $\bm{\Theta}_{\text{OP}}=[\alpha, \beta, \eta]$ and $\bm{\Theta}_{\text{MO}}=[c, \mu, \eta]$, respectively for optical and molecular channels. In this work, we assume that the sampling rate $\omega$, and the proportionality constants $\kappa_{\text{OP}}$ and $\kappa_{\text{MO}}$ are fixed and are not part of the model parameters. Note that $\alpha$ and $\beta$ can change over time due to atmospheric turbulence or mobility. Similarly, $c$ and $\mu$ are functions of the distance between the transmitter and the receiver, flow velocity, and the diffusion coefficient, which may change over time, e.g., due to variations in temperature and pressure \cite{far16ST}. The background noise $\eta$ may also change with time. Note that although the symbol interval $\tau$ may be changed to increase or decrease the data rate, both the transmitter and receiver must agree on the value of $\tau$. Thus, we assume that the value of $\tau$ is always known at the receiver, and therefore, it is not part of the CSI.  %Although the marginal distribution of the \jth~sample of the \kth~symbol are Poisson distributed according to \eqref{eq:PoissonMarginal}, it is very difficult to obtain the joint distribution in \eqref{eq:Poisson}, or to numerically calculate it \cite{kar05,ino17}. However, 
In the next subsection, we present the optimal VD, assuming that the receiver knows all the model parameters $\bm{\Theta}_{\text{OP}}$ and $\bm{\Theta}_{\text{MO}}$ perfectly.

\subsection{The Viterbi Detector}
\label{sec:ViterbiDetect}
The VD assumes a certain memory length $M$ where the current observed signal is affected only by the past $M$ transmitted symbols. In this case \eqref{eq:PoissonMarginal} becomes 
\begin{align}
	\label{eq:PoissonMemMarg}
	\vec{y}_k[j] \sim \PoisDist\left(x_k\bm{\lambda}_{0}[j] + \sum_{l=1}^{M} x_{k-l} \bm{\lambda}_{l}[j] + \eta \right). 
\end{align}
Since the marginal distribution of the \jth~sample of the \kth~symbol is Poisson distributed according to \eqref{eq:PoissonMemMarg}, given the model parameters $\bm{\Theta}_{\text{pois}}$, we have
\begin{align}
	\footnotesize
	\label{eq:assumption}
	P(\vec{y}_k \mid x_{k-M}, &x_{k-M+1},\cdots,x_k,\bm{\Theta}_{\text{pois}}) = \nonumber \\
	&\prod_{j=1}^a P(\vec{y}_k[j] \mid x_{k-M}, x_{k-M+1}, \cdots, x_k, \bm{\Theta}_{\text{pois}}).	
\end{align}
This is because, given the model parameters as well as the current symbol and the previous $M$ symbols, the samples within the current bit interval are generated independently and distributed according to \eqref{eq:PoissonMemMarg}. Note that \eqref{eq:assumption} holds only if the memory length $M$ is known perfectly. If the estimate of $M$ is inaccurate, then \eqref{eq:assumption} is also inaccurate.

Let $\mathcal{V}=\{v_0,v_1,\cdots,v_{2^M-1}\}$ be the set of states in the trellis of the VD, where the state $v_u$ corresponds to the previous $M$ transmitted bits $[x_{-M},x_{-M+1},\cdots,x_{-1}]$ forming the binary representation of $u$. Let $\hat{x}_k$, $1\leq k \leq K$ be the information bits to be estimated. Let $V_{k,u}$ be the state corresponding to the \kth~symbol interval, where $u$ is the binary representation of  $[\hat{x}_{k-M},\hat{x}_{k-M+1},\cdots,\hat{x}_{k-1}]$. Let $\mathcal{L}(V_{k,u})$ denote the log-likelihood of the state $V_{k,u}$. For a state $V_{k+1,u} = [\hat{x}_{k-M+1}, \hat{x}_{k-M+2}, \cdots, \hat{x}_k]$, there are two states in the set $\{V_{k,i}\}_{i=0}^{2^M-1}$ that can transition to $V_{k+1,u}$:
\begin{align}
	u_0& = \lfloor \tfrac{u}{2} \rfloor, \\
	u_1& = \lfloor \tfrac{u}{2} \rfloor + 2^{M-1},
\end{align}
where $\lfloor . \rfloor$ is the floor function. Let the binary vector $\vec{b}_{u_0} = [0, \hat{x}_{k-M+1}, \hat{x}_{k-M+2}, \cdots, \hat{x}_{k-1}]$ be the binary representation of $u_0$ and similarly $\vec{b}_{u_1}$ the binary representation of $u_1$. The log-likelihoods of each state in the next symbol slot are updated according to
\begin{align}
	\mathcal{L}(V_{k+1,u}) = \max[\mathcal{L}(V_{k,u_0}) &+\mathcal{L}(V_{k,u_0},V_{k+1,u}), \nonumber \\
									\mathcal{L}(V_{k,u_1}) &+\mathcal{L}(V_{k,u_1},V_{k+1,u})],
\end{align} 
where $\mathcal{L}(V_{k,u_i},V_{k+1,u})$, $i\in\{0,1\}$, is the log-likelihood increment of transitioning from state $V_{k,u_i}$ to $V_{k+1,u}$. Let
\begin{align}
\label{eq:bigLambda}
\bm{\Lambda}_{u_i,u}[j]\mspace{-3mu}=\mspace{-3mu}(u \mspace{-3mu}\mod 2) \bm{\lambda}_{0}[j]\mspace{-3mu} +\mspace{-3mu} \sum_{l=1}^{M} \vec{b}_{u_i}[M\mspace{-3mu}-\mspace{-3mu}l\mspace{-3mu}+\mspace{-3mu}1] \bm{\lambda}_{l}[j] \mspace{-3mu}+ \mspace{-3mu}\eta.
\end{align} 
Using the PMF of the Poisson distribution, \eqref{eq:PoissonMemMarg}, \eqref{eq:assumption}, and \eqref{eq:bigLambda} we have
\begin{align}
\mathcal{L}(V_{k,u_i},V_{k+1,u})\mspace{-3mu} = \mspace{-3mu}-\mspace{-3mu}\sum_{j=1}^a \bm{\Lambda}_{u_i,u}[j]\mspace{-3mu} +\mspace{-3mu}\sum_{j=1}^a \log(\bm{\Lambda}_{u_i,u}[j]) \vec{y}_k[j],
\end{align}  
where the extra term $-\sum_{j=1}^a \log(\vec{y}_k[j]!)$ is dropped since it will be the same for both transitions from $u_0$ and $u_1$. Using these transition probabilities and setting the $\mathcal{L}(V_{0,0})=0$ and $\mathcal{L}(V_{0,u})=-\infty$, for $u\neq0$, the most likely sequence $\hat{x}_k$, $1\leq k \leq K$, can be estimated using the Viterbi algorithm~\cite{for73}. When the memory length is long, it is not computationally feasible to consider all the states in the trellis as they grow exponentially with memory length. Therefore, in this work we implement the Viterbi beam search algorithm \cite{lin04VDbeam}. In this scheme, at each time slot, only the transition from the previous $N$ states with the largest log-likelihoods are considered. When $N=2^M$, the Viterbi beam search algorithm reduces to the traditional Viterbi algorithm.

We now evaluate the performance of NN detectors using the Poisson channel model.

%\begin{figure*}[t!]
%	% ensure that we have normalsize text
%	\normalsize
%	%\captionsetup{font=footnotesize}
%	\centering
%	% ensure that we have normalsize text
%	\begin{minipage}{.43\textwidth}
%		\begin{center}
%			\includegraphics[width=\columnwidth,keepaspectratio]{TPAMI_Optical_BER_vs_TopX}
%			(a)
%		\end{center}	
%	\end{minipage}
%	%\vspace{-0.1cm}
%	\hspace{0.8cm}
%	\begin{minipage}{.43\textwidth}
%		\begin{center}
%			\includegraphics[width=\columnwidth,keepaspectratio]{TPAMI_Molecular_BER_vs_TopX}
%			(b)
%		\end{center}
%	\end{minipage}
%	\caption{\label{fig:BERvsTopX} Performance of the VD beam search as function of $N$. (a) Optical channel with $\bm{\Theta}_{\text{OP}}=[\beta=0.2,\eta=1]$ and $\tau=0.025$ $\mu$s. (b) Molecular channel with $\bm{\Theta}_{\text{MO}}=[c=8,\mu=40,\eta=100]$ and $\tau=0.5$ s.  }
%\end{figure*}

\section{Evaluation Based on Poisson Channel}
\label{sec:resultPoiss}
In this section we evaluate the performance of the proposed SBRNN detector based on the Poisson channel model, and in the next section we use the experimental platform developed in \cite{far17ExpPlat} to demonstrate that the SBRNN detector can be implemented in practice to perform real-time detection. The rest of this section is organized as follows. First, we describe the training procedure and the simulation setup in Section \ref{subsec:simsetup}. Then, in Section \ref{subsec:SLSDN}, we evaluate the effects of $L_{\max}$ and $M$, the symbol duration, and noise on the BER performance. In particular, in this section we demonstrate that SBRNN detection is resilient to changes in symbol duration and noise, and outperforms VD with perfect CSI if the memory length $M$ is not estimated correctly.  In Section \ref{subsec:ChanParams}, the performance of the SBRNN detector and VD are evaluated for different channel parameters. To show that the SBRNN algorithm works on larger symbol sets (i.e., higher order modulations), in section \ref{subsec:sybolset} we consider an optical channel that uses $m$-PAM, $m>2$, instead of OOK (i.e., 2-PAM). We also demonstrate that although the training is performed on transmission sequences of length 100, the SBRNN can generalize to longer transmission sequences. The effects of  the RNN cell type is also evaluated and it is demonstrated that LSTM cells achieve the best BER performance.  The performance of the SBRNN in rapidly changing channels is evaluated in Section \ref{subsec:changChan}, and the complexity of this algorithm compared to the VD is discussed in Section \ref{subsec:CompComplex}. Table \ref{tb:resultsSummary} summarizes all the results that will be presented in this section.

\begin{table}[h]
	\scriptsize 
	\caption{Summary of the results to be presented in this section.}
	\label{tb:resultsSummary}
	\centering
	\begin{tabular}{c|lll}
		\toprule
		sec.		& Chan Types  		& Evaluates \\
		\midrule
		B							& Optical/Molecular (OOK) 	& sequence length, symbol duration, noise\\
		C		& Optical/Molecular (OOK)      & channel parameters (i.e., impulse response)\\
		D		& Optical ($m$-PAM)  & symbol size, transmission length, RNN type\\
		E	& Optical/Molecular	(OOK) & rapidly changing channels\\
		\bottomrule
	\end{tabular}
	\vspace{-0.5cm}
\end{table}
%   

%to compare the performance of the SBRNN to an RNN detector, the VD detector with perfect CSI estimates, and the VD with imperfect CSI estimates. We demonstrate that unlike the RNN detector, the SBRNN detector is robust to changing channel conditions and can outperform the VD with CSI estimation error. Second, we use the experimental platform developed in \cite{far17ExpPlat} to demonstrate that the SBRNN detector, which does not require any underlying analytical channel model, can be implemented in practice to perform real-time detection. In particular, we use the experimental data collected on the platform to train and implement the SBRNN as part of a molecular communication system for real-time text messaging. We show that SBRNN outperforms all the other NN detectors we consider in terms of BER performance. 

%
\begin{figure*}
	\centering
	\includegraphics[width=1\textwidth,keepaspectratio]{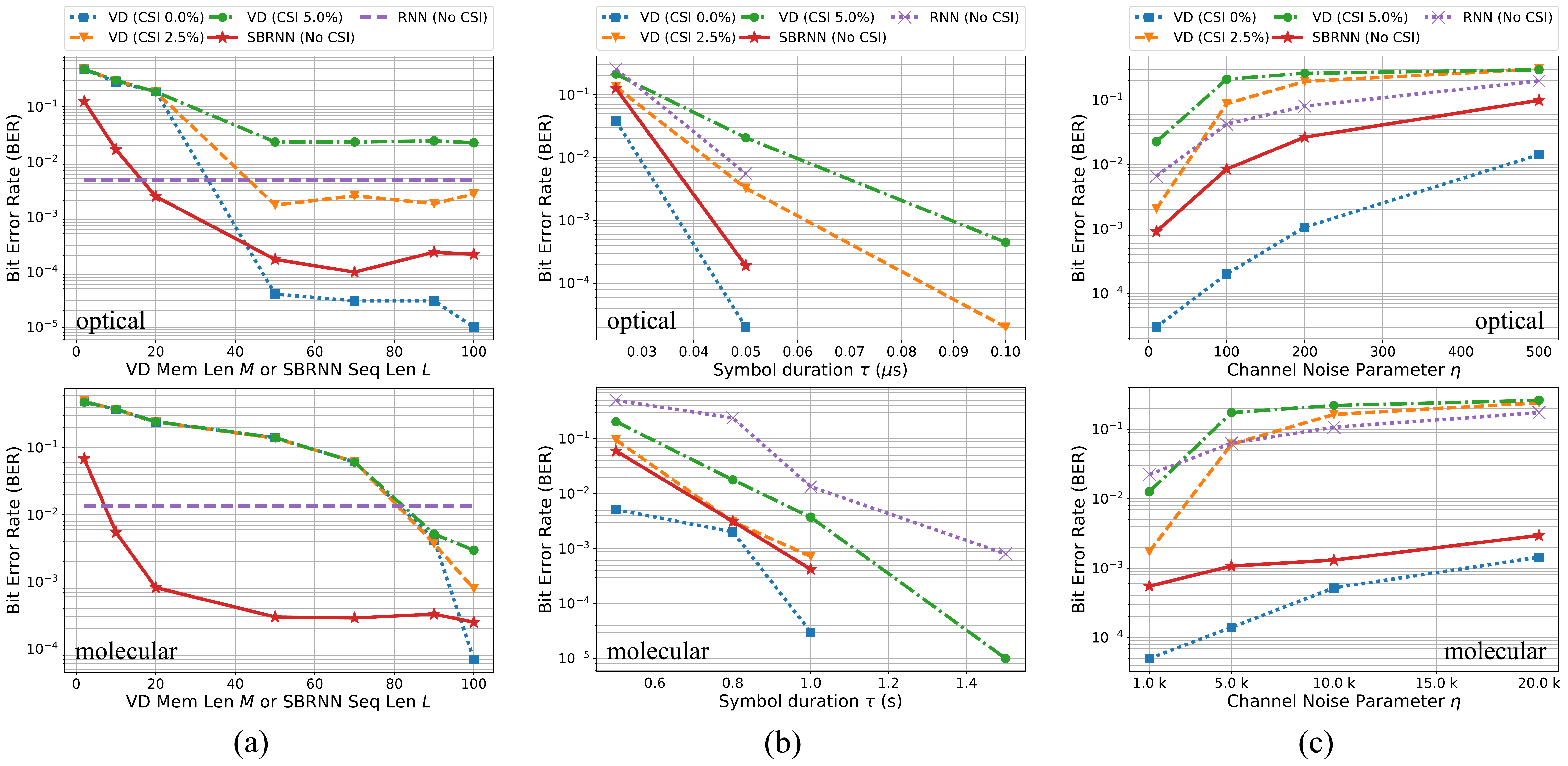}
	%\vspace{-0.3cm}
	\caption{\label{fig:poisBERvSeqTimeEta} The BER performance comparison of the SBRNN detector, the RNN detector, and the VD. The top plots present the optical channel and the bottom plots present the molecular channel. (a) The BER at various memory lengths $M$ and SBRNN sequence lengths $L$. Top: $\bm{\Theta}_{\text{OP}}=[\beta=0.2,\eta=1]$ and $\tau=0.05$ $\mu$s. Bottom:  $\bm{\Theta}_{\text{MO}}=[c=10,\mu=40,\eta=100]$ and $\tau=1$ s. (b) The BER at various symbol durations for $L=50$ and $M=99$. The $\bm{\Theta}_{\text{OP}}$ (top) and $\bm{\Theta}_{\text{MO}}$ (bottom) are the same as (a). (c) The BER at various noise rates for $L=50$ and $M=99$. Except $\eta$, all the parameters are the same as those in (a). }
	\vspace{-0.4cm}
\end{figure*}

\subsection{Training and Simulation Procedure}
\label{subsec:simsetup}

For evaluating the performance of the SBRNN on the Poisson channel, we consider both the optical channel and the molecular channel. For the optical channel, we assume that the channel parameters are $\bm{\Theta}_{\text{OP}}=[\beta,\eta]$, and assume $\alpha=2$ and $\kappa_{\text{OP}}=10$. We use these values for $\alpha$ and $\kappa_{\text{OP}}$ since they resulted in system responses that resembled the ones presented in \cite{hay07OptiGamma, maj07OptiGamma, din09OptiGamma}. For the molecular channel the model parameters are $\bm{\Theta}_{\text{MO}}=[c,\mu,\eta]$, and $\kappa_{\text{MO}}=10^4$. The value of $\kappa_{\text{MO}}$ was selected to resemble the system response in \cite{far13}. For the optical channel we use $\omega=2$ GS/s and for the molecular channel we use $\omega=100$ S/s.

For the VD algorithm we consider Viterbi with beam search, where only the top $N=100$ states with the largest log-likelihoods are kept in the trellis during each time slot. We also consider two different scenarios for CSI estimation. In the first scenario we assume that the detector estimates the CSI perfectly, i.e., the values of the model parameters $\bm{\Theta}_{\text{OP}}$ and $\bm{\Theta}_{\text{MO}}$ are known perfectly at the receiver. In practice, it may not be possible to achieve perfect CSI estimation. In the second scenario we consider the VD with CSI estimation error. Let $\zeta$ be a parameter in $\bm{\Theta}_{\text{OP}}$ or $\bm{\Theta}_{\text{MO}}$. Then the estimate of this parameter is simulated by  $\hat{\zeta}=\zeta+Z$, where $Z$ is a zero-mean Gaussian noise with a standard deviation that is 2.5\% or 5\% of $\zeta$. In the rest of this section, we refer to these cases as the VD with 2.5\% and 5\% error, and the case with perfect CSI as the VD with 0\% error. Table~\ref{tb:BERvsTopX} shows the BER performance of the VD for different values of $N$. It can be seen that $N=100$, which is used in the rest of this section, is sufficient to achieve good performance with the VD.
\begin{table}[t]
	\scriptsize 
	\caption{Performance of the VD beam search as function of $N$. The optical channel results is obtained using $\bm{\Theta}_{\text{OP}}=[\beta=0.2,\eta=1]$ and $\tau=0.025$ $\mu$s and the molecular channel results using $\bm{\Theta}_{\text{MO}}=[c=8,\mu=40,\eta=100]$ and $\tau=0.5$ s. }
	\label{tb:BERvsTopX}
	\centering
	\begin{tabular}{c|lllll}
		\toprule
		$N$ 						& 10  		& 100 		& 200 		& 500 		& 1000  \\
		\midrule
		Opti. VD 0.0\% error		& 0.0466  	& 0.03937	& 0.03972  	& 0.03906   & 0.03972\\
		Opti. VD 2.5\% error		& 0.226     & 0.175     & 0.17561  	& 0.15889   & 0.1509 \\
		Opti. VD 5.0\% error		& 0.4036    & 0.385     & 0.38519   & 0.39538  	& 0.36\\
		Mole. VD 0.0\% error	& 0.00466	& 0.00398  	& 0.00464 	& 0.00448  	& 0.00432\\
		Mole. VD 2.5\% error	& 0.0066	& 0.0055  	& 0.00524 	& 0.0056  	& 0.00582\\
		Mole. VD 5.0\% error	& 0.41792	& 0.34667   & 0.30424 	& 0.29314  	& 0.30588\\
		\bottomrule
	\end{tabular}
	\vspace{-0.5cm}
\end{table}

Both the RNN and the SBRNN detectors use LSTM cells \cite{hoc97}, unless specified otherwise. For the SBRNN, the size of the output is 80. For the RNN, since the SBRNN uses two RNNs, one for the forward direction and one for the backward direction, the size of the output is 160. This ensures that the SBRNN detector and the RNN detector have roughly the same number of parameters. The number of layers used for both detectors in this section is 3. The input to the NNs are a set of normalized features $\vec{r}_k$ extracted from the received signal $\vec{y}_k$. The feature extraction algorithm is described in the appendix. This feature extraction step normalizes the input, which assists the NNs to learn faster from the data \cite{goodfellowBook}.

To train the RNN and SBRNN detectors, transmitted bits are generated at random and the corresponding received signal is generated using the Poisson model in \eqref{eq:PoissonMarginal}. In particular, the training data consists of many samples of sequences of 100 consecutively transmitted bits and the corresponding received signal. Since in this work we focus on uncoded communication, we assume the occurrence of both bits in the transmitted sequence are equiprobable. For each sequence, the CSI are selected at random. Particularly, for the optical channel, for each 100-bit sequence,
\begin{align}
\beta &\sim \UniDist(\{0.15,0.16,0.17,\cdots,0.35\}), \nonumber \\ 
\eta  &\sim\UniDist(\{1, 10, 20, 50, 100, 200, 500\}), \nonumber \\ 
\tau &\sim \UniDist(\{0.025, 0.05, 0.075, 0.1\}) \text{(all in $\mu$s)}, \label{eq:opticalParamRange}
\end{align}
where $\UniDist(\mathcal{A})$ indicates uniform distribution over the set $\mathcal{A}$. Similarly, for the molecular channel,
\begin{align}
c  &\sim \UniDist(\{1,2,\cdots,30\}), \nonumber \\ 
\mu &\sim \UniDist(\{5,10,15,\cdots,65\}), \nonumber \\
\eta&\sim\UniDist(\{1, 50, 100, 500, 1\text{k}, 5\text{k}, 10\text{k}, 20\text{k}, 30\text{k}, 40\text{k}, 50\text{k}\}), \nonumber \\
\tau&\sim \UniDist(\{0.5, 1, 1.5, 2\}) \text{(all in s)}. \label{eq:molecularParamRange}
\end{align}
  %$\beta \sim \UniDist(\{0.15,0.16,0.17,\cdots,0.3\})$, $\eta\sim\UniDist(\{1, 10, 20, 50, 100, 200, 500\})$, and $\tau\sim \UniDist(\{0.025, 0.05, 0.075, 0.1\})$ (all in $\mu$s), where $\UniDist(\mathcal{A})$ indicates uniform distribution over the set $\mathcal{A}$. For molecular channel, for each 100-bit sequence, $c \sim \UniDist(\{1,2,\cdots,30\})$, $\mu\sim \UniDist(\{5,10,15,\cdots,65\})$, $\eta\sim\UniDist(\{1, 50, 100, 500, 1\text{k}, 5\text{k}, 10\text{k}, 20\text{k}, 30\text{k}, 40\text{k}, 50\text{k}\})$, and $\tau\sim \UniDist(\{0.5, 1, 1.5, 2\})$ (all in seconds). 
For the SBRNN training, each 100-bit sequence is randomly broken into subsequences of length $L\sim \UniDist(\{2,3,4,\cdots,50\})$. For all training, the Adam optimization algorithm \cite{kin14} is used with learning rate of $10^{-3}$, and batch size of 500. We train on 500k sequences of 100 bits.

Over the next several subsections we evaluate the performance of the SBRNN detector and compare it to that of the VD.

\begin{figure*}
	\centering
	\includegraphics[width=1\textwidth,keepaspectratio]{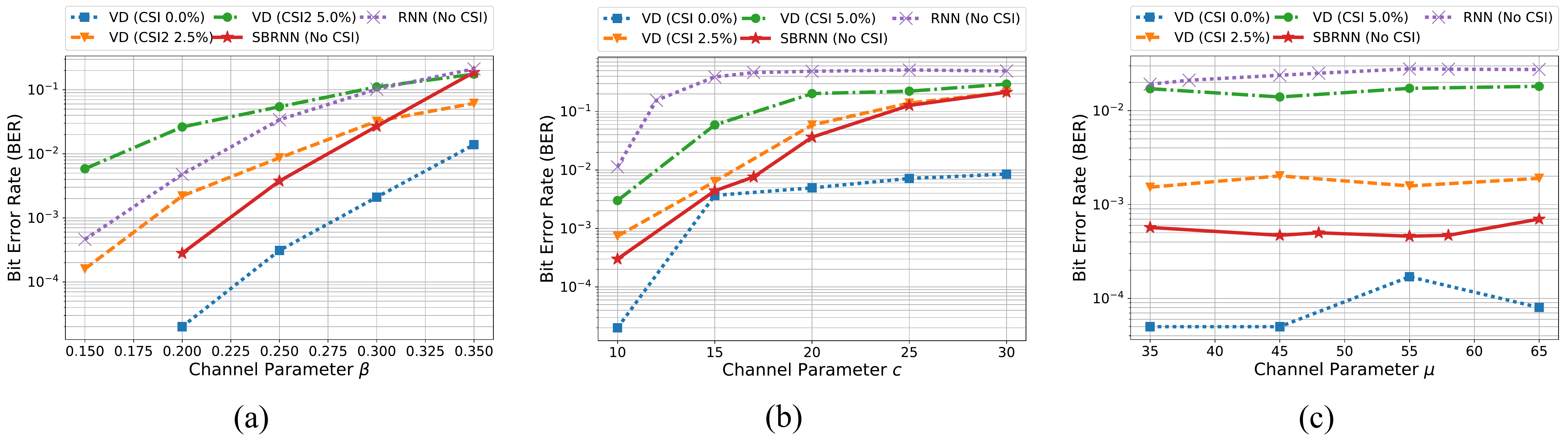}
	%\vspace{-0.3cm}
	\caption{\label{fig:poisBERvBetaMuC} The BER performance comparison of the SBRNN detector ($L=50$), the RNN detector, and the VD ($M=99$). (a)~The BER at various $\beta$ for the optical channel with $\eta=1$ and $\tau=0.05$ $\mu$s. (b)~The BER at various $c$ for the molecular channel with $\mu=40$, $\eta=1$, and $\tau=1$ s. (c)~The BER at various $\mu$ for the molecular channel with $c=10$, $\eta=1000$, and $\tau=1$ s.}
	\vspace{-0.4cm}
\end{figure*}

\subsection{Effects of Sequence Length, Symbol Duration, and Noise}
\label{subsec:SLSDN}
First, we evaluate the BER performance with respect to the memory length $M$ used in the VD, and the sequence length $L$ used in the SBRNN. For all the BER performance plots in this section, to calculate the BER, 1000 sequences of 100 random bits are used.  Figure~\ref{fig:poisBERvSeqTimeEta}(a) shows the results for the optical (top plots) and the molecular (bottom plots) channels with the parameters described above. From the results it is clear that the performance of the VD relies heavily on estimating the memory length of the system correctly. We define the memory length as the number of symbol durations it takes for the impulse response to be {\em sufficiently small} such that ISI is negligible or, equivalently, such that increasing the memory length of the detector does not decrease BER significantly.  For example, let $\lambda_{\max}$ be the peak value of the impulse response. Let $t_{\sigma}$, $0<\sigma<1$, be the time it takes for impulse response to fall to $\sigma \lambda_{\max}$. Then, for the optical channel in Figure~\ref{fig:poisBERvSeqTimeEta}(a), the time it takes for the impulse response to fall to 0.01\% of $\lambda_{\max}$ is $\tau_{0.0001}=2.55$ $\mu$s. Therefore, at a symbol duration of $\tau=0.05$, the memory length is on the order of $M\approx51$ symbols. From Figure~\ref{fig:poisBERvSeqTimeEta}(a) it can be seen that the BER performance of the VD with perfect CSI does not improve beyond a negligible amount for $M>50$. The molecular channel's impulse response has a much longer tail, where at $\tau=1$ s it takes 382 symbol durations for the impulse response to fall to 0.1\% of the peak value $\lambda_{\max}$. This is evident in  Figure~\ref{fig:poisBERvSeqTimeEta}(a) where the BER of the VD with perfect CSI always improves as $M$ increases. 

 Figure~\ref{fig:poisBERvSeqTimeEta}(a) also demonstrates that if the estimate of $M$ is inaccurate, the SBRNN algorithm outperforms the VD with perfect CSI. We also observe that the SBRNN achieves a better BER when there is a CSI estimation error of 2.5\% or more. Note that the RNN detector does not have a parameter that depends on the memory length and has a significantly larger BER compared to the SBRNN. For the optical channel, the RNN detector outperforms the VD with 5\% error in CSI estimation. Moreover, it can seen that the optical channel has a shorter memory length compared to the molecular channel. 

{\bf \em Remark 1:} When the VD has perfect CSI, it can estimate the memory length correctly by using the system response. However, if there is CSI estimation error, the memory length may not be estimated correctly, and as can be seen in Figure~\ref{fig:poisBERvSeqTimeEta}(a), this can have degrading effects on the performance of the VD. However, in the rest of this section, for all the other VD plots, we use the memory length of 99, i.e., the largest possible memory length in sequences of 100 bits. Although this does not capture the performance degradation that may result from the error in estimating the memory length, as we will show, the SBRNN still achieves a BER performance that is as good or better than the VD plots with CSI estimation error under various channel conditions.

Next we evaluate the BER for different symbol durations in Figure~\ref{fig:poisBERvSeqTimeEta}(b). Again we observe that the SBRNN achieves a better BER when there is a CSI estimation error of 2.5\% or more. The RNN detector outperforms the VD with 5\% CSI estimation error for the optical channel, but does not perform well for the molecular channel. All detectors achieve zero-error in decoding the 1000 sequences of 100 random bits used to calculate the BER for the optical channel with $\tau=0.1$ $\mu$s. Similarly, for the molecular channel at $\tau=1.5$ s, all detectors except the RNN detector achieve zero error.   

Figure~\ref{fig:poisBERvSeqTimeEta}(c) evaluates the BER performance at various noise rates. The SBRNN achieves a BER performance close to the VD with perfect CSI across a wide range of values. For larger values of $\eta$, i.e., low signal-to-noise ratio (SNR), both the RNN detector and the SBRNN detector outperform the VD with CSI estimation error.

\begin{figure}
	\centering
	\includegraphics[width=1\columnwidth,keepaspectratio]{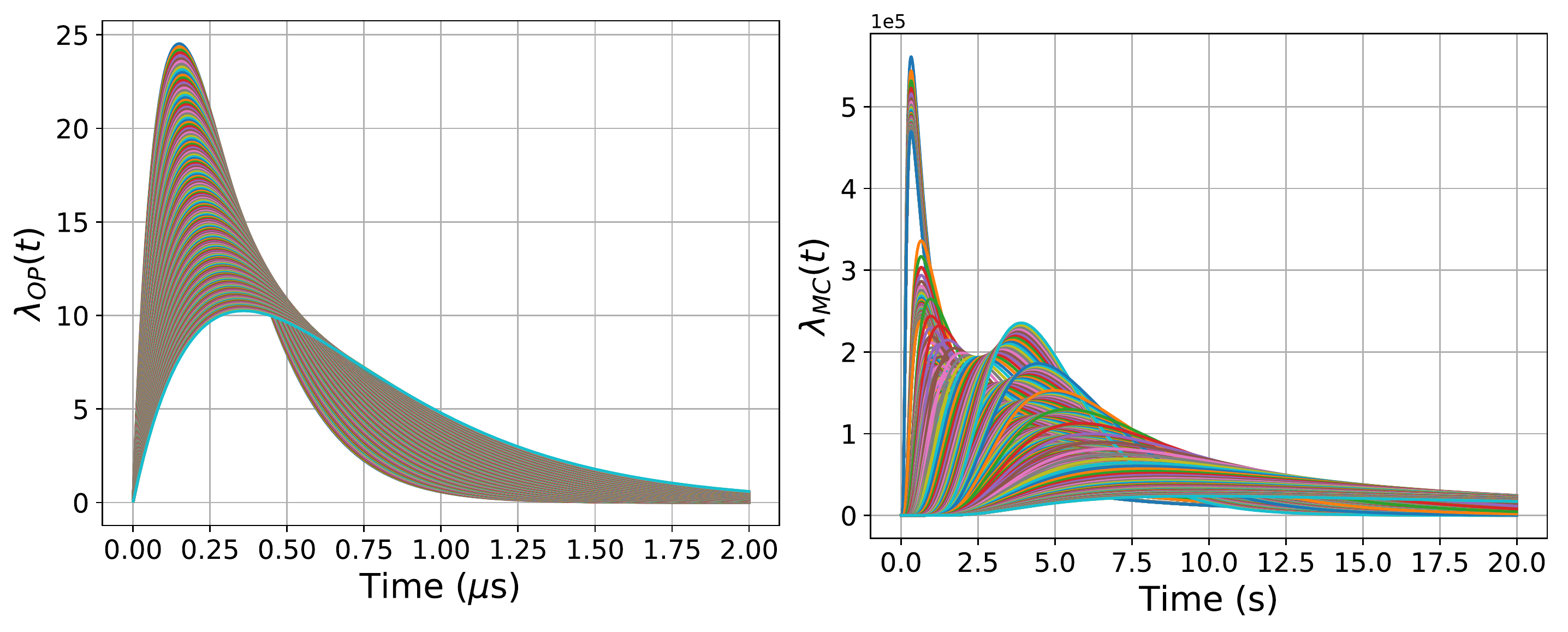}
	%\vspace{-0.3cm}
	\caption{\label{fig:responseRange} The shape the system response for the optical and molecular channel over the range of values in \eqref{eq:opticalParamRange} and \eqref{eq:molecularParamRange}.}
	\vspace{-0.6cm}
\end{figure}

\subsection{Effects of Channel Parameters}
\label{subsec:ChanParams}

In this section we evaluate the performance with respect to the channel parameters that affect the system response. Recall that for the optical channel the parameter $\beta$ affects the system response in \eqref{eq:impulseResOpti} (note that here we assume that $\alpha=2$ does not change), and for the molecular channel the parameters $c$ and $\mu$ affect the system response in \eqref{eq:impulseRes}. The range of values that $\beta$ is assumed to take is given in \eqref{eq:opticalParamRange}, and the range of values for $c$ and $\mu$ are given in \eqref{eq:molecularParamRange}.

In Figure~\ref{fig:poisBERvBetaMuC}, we evaluate the performance of the detection algorithms with respect to these parameters. Note that in optical and molecular communication these parameters can change rapidly due to atmospheric turbulence, changes in temperature, or changes in the distance between the transmitter and the receiver. Therefore, estimating these parameters accurately can be challenging. Furthermore, since these parameters change the shape of the system responses they change the memory length as well. 

Figure \ref{fig:responseRange} shows the system response for the optical and molecular channels over the range of values for $\beta$, $c$, and $\mu$ in \eqref{eq:opticalParamRange} and \eqref{eq:molecularParamRange}. For a fixed symbol duration, the system response can have a considerable effect on the delay spread (i.e., memory order) of the system. From Figure~\ref{fig:poisBERvBetaMuC}, it can be seen that the SBRNN performs as well or better than the VD with an estimation error of  2.5\%. Moreover, for the optical channel, the RNN detector performs better than the VD with 5\% estimation error.   In all cases, the SBRNN learns to detect over the wide range of system responses shown in Figure \ref{fig:responseRange}.

%%
%\begin{figure}
%	\centering
%	\includegraphics[width=0.8\columnwidth,keepaspectratio]{TPAMI_BarPlot_ModulationComparison.pdf}
%	%\vspace{-0.3cm}
%	\caption{\label{fig:modCompare} \revised{The performance comparison of OOK, 4-PAM, and 8-PAM.}}
%	
%\end{figure}
%%

%
\begin{figure*}
	\centering
	\includegraphics[width=1\textwidth,keepaspectratio]{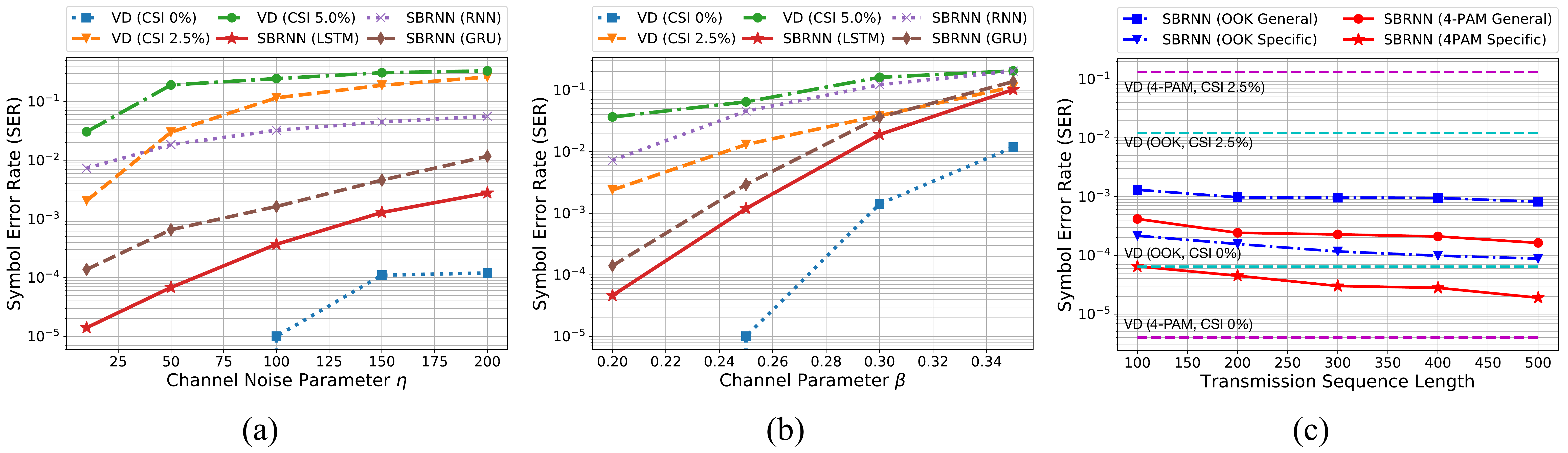}
	%\vspace{-0.3cm}
	\caption{\label{fig:4PAMTranLen} The SER performance comparison of the SBRNN detector ($L=50$), and the VD ($M=99$) for optical channel with 4-PAM modulation. (a)~The SER at various $\eta$ for the optical channel with $\beta=0.2$ and $\tau=1$ $\mu$s. (b)~The SER at various $\beta$ for the optical channel with $\eta=10$ and $\tau=1$ $\mu$s. (c)~The SER versus transmission sequence length for optical channel with OOK modulation ($\tau=0.05$ $\mu$s, $\kappa_{\text{OP}}=10$, $\eta=50$) and 4-PAM modulation ($\tau=0.1$ $\mu$s, $\kappa_{\text{OP}}=20$, $\eta=100$).  For both cases $\beta = 0.2$ and $\alpha=2$.}
	\vspace{-0.6cm}
\end{figure*}

\subsection{Effects of Symbol Set Size, Transmission Length, and RNN Cell Type}
\label{subsec:sybolset}

\begin{table}[t]
	\scriptsize 
	\caption{The SER for different modulations.}
	\vspace{-0.3cm}
	\label{tb:SER_mPAM}
	\centering
	\begin{tabular}{c|cc}
		\toprule
		Modulation			& SER (Perfect CSI) 		& SER (2.5\% CSI Error) \\
		\midrule
		OOK				       &$6.4\times 10^{-5}$		& $9.2\times 10^{-3}$  \\
		4-PAM   			&$5.3\times 10^{-6}$			& $1.3\times 10^{-1}$  \\
		8-PAM    			&$1.6\times 10^{-4}$	    	& $2.5\times 10^{-1}$  \\
		\bottomrule
	\end{tabular}
	\vspace{-0.4cm}
\end{table}

In the previous sections we considered OOK modulation. However, it is not clear if higher order modulations can be used to achieve better results. In this section we first evaluate the performance of OOK and higher order $m$-PAM modulations using VD. We demonstrate that for system parameters under consideration, 4-PAM achieves the best BER performance. Then we demonstrate that the SBRNN detector can be trained on modulations with larger symbol sets. In fact for detection and estimation problems in speech and language processing, where RNNs are extensively used, the symbol set (i.e., the number of phonemes or vocabulary size) can be on the order of hundreds to millions of symbols. We also consider the affect of different RNN cell types on the symbol error rate (SER) performance and demonstrate that the LSTM cell, which was used in the previous sections, achieves the best performance.  Finally, the generalizability of the SBRNN detector to longer transmission sequences is evaluated where we show that the SBRNN achieves the same or better SER performance on longer transmission sequences, despite being trained on sequences of length 100.

    	First we compare the performance of OOK, 4-PAM, and 8-PAM modulation, where 2, 4 or 8 amplitude levels are used for encoding 1, 2 or 3 bits of information during each symbol duration. We assume that amplitudes are equally spaced and include the zero amplitude (i.e., sending no pulse). Because of space limitations, we only focus on the optical channel with the following parameters: OOK with $\tau=0.05$ $\mu$s, $\kappa_{\text{OP}}=10$, $\eta=50$; 4-PAM with $\tau=0.1$ $\mu$s, $\kappa_{\text{OP}}=20$, $\eta=100$; and 8-PAM with $\tau=0.15$ $\mu$s, $\kappa_{\text{OP}}=30$, $\eta=150$.  For all modulations we use $\beta=0.2$ and $\alpha=2$. We chose these parameters to keep the average transmit power, the data rate, and the peak signal-to-noise ratio (SNR) the same for all modulations. We then evaluate the SER using the VD  with perfect CSI and the VDs with  CSI estimation errors of 2.5\%.  We use 500k symbols for evaluating the SER. Table~\ref{tb:SER_mPAM} shows the results. When perfect CSI is available at the receiver, 4-PAM achieves the best SER, while when there is an error in CSI estimation, OOK achieves the best SER. Note that since the number of bits presented by each symbol of each  modulation scheme is different, SER is not the best performance measure. However, even if we assume that each symbol error is due to {\em a single} bit error, which results in the best BER possible for 8-PAM,  we still observe that 4-PAM achieves the best BER performance when perfect CSI is available at the receiver, while OOK achieves the best BER performance when there is CSI estimation error.

Since 4-PAM achieves the best BER performance, we trained a new SBRNN detector based on 4-PAM modulation. For training, the channel parameter $\beta$ is assumed to be uniformly random in the interval $\beta\in[0.2, 0.35]$ and the noise parameter $\eta$ is assumed to be uniformly random in the interval $\eta\in[10, 200]$. We trained three SBRNN detectors based on the LSTM cell, the GRU cell~\cite{cho2014learning},  and the vanilla RNN architecture~\cite{goodfellowBook}. Figure~\ref{fig:4PAMTranLen}(a)-(b) shows the results.  As can be seen, the SBRNN with the LSTM cell achieves a better SER performance compared with the GRU cell and the vanilla RNN cell types. Compared with the VDs, we observe a trend similar to that in OOK modulation: the SBRNN outperforms VD with CSI estimation error, while its performance comes close to the VD with perfect CSI. This demonstrates that the SBRNN algorithm can be extended to larger symbol sets. 

		We last evaluate the performance of the SBRNN detector over longer transmission sequences for OOK and 4-PAM. In particular, for each modulation, two differently trained SBRNN networks are evaluated. The first set of networks are the  same networks used to generate Figures~\ref{fig:poisBERvSeqTimeEta}, \ref{fig:poisBERvBetaMuC}, and \ref{fig:4PAMTranLen}(a)-(b). These networks are trained using a data set that contains sample transmissions under various channel conditions. The second set of networks are trained using sample received signals from a very specific set of channel and noise parameters. Specifically, the training data is generated using the same set of parameters that are used during testing (i.e., $\tau=0.05$ $\mu$s, $\kappa_{\text{OP}}=10$, $\eta=50$ for OOK and $\tau=0.1$ $\mu$s, $\kappa_{\text{OP}}=20$, $\eta=100$ for 4-PAM). Note that all the SBRNN detectors are trained on transmission sequences of length 100. 
		
\begin{figure}
	\centering
	\includegraphics[width=1\columnwidth,keepaspectratio]{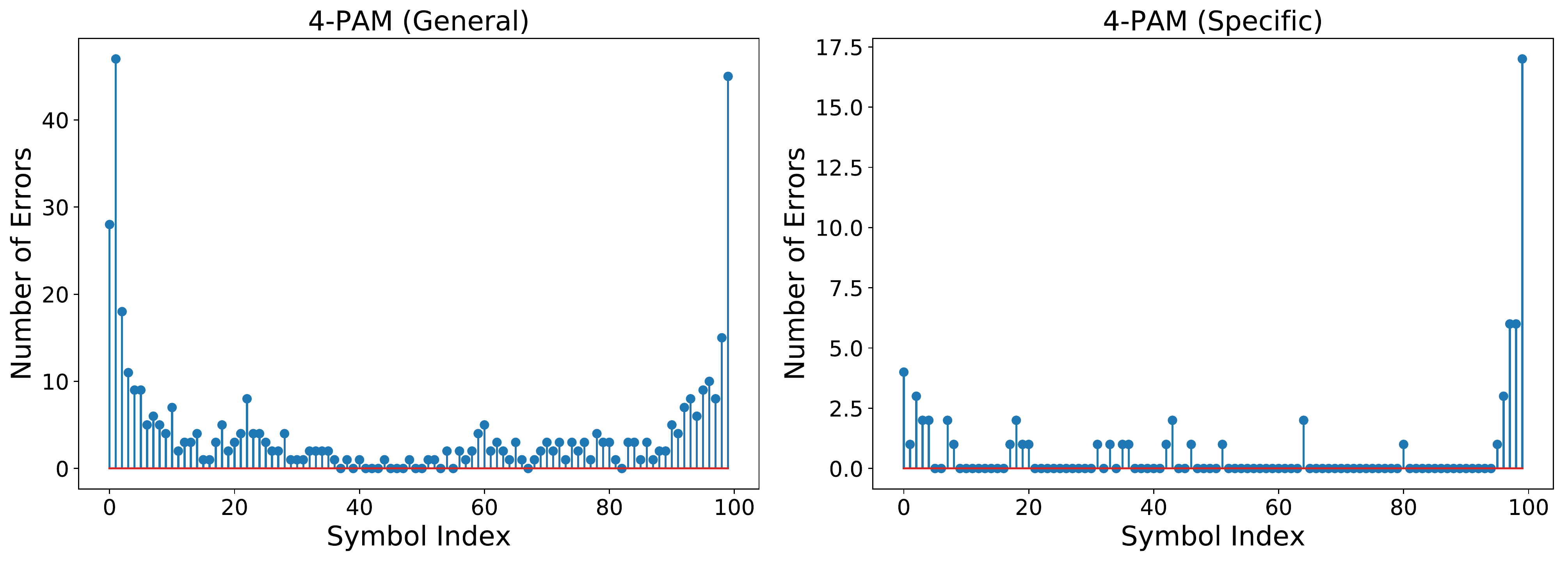}
	\vspace{-0.3cm}
	\caption{\label{fig:errorPositions} Symbols errors are higher at the beginning and end of the transmission sequence.}
	\vspace{-0.5cm}
\end{figure}

		Figure~\ref{fig:4PAMTranLen}(c) shows the performance for transmission sequences of various lengths. Interestingly, we observe that the SER drops as the length of the transmission sequence increases. This is because the probability of error for symbols at the beginning and end of the  transmission sequence is higher as shown in Figure~\ref{fig:errorPositions}. The larger probability of error for the first few symbols is due to the signal rising rapidly at the start of the transmission, as was shown in Figure~6, which has a different structure compared to the signal corresponding to the rest of the symbols. This can be mitigated by using a separate neural network that is trained only on the signal corresponding to the initial symbols, or using a sequence of random transmission bits at the beginning of the transmission sequence as a guard interval. The error at the end of transmission sequence can be mitigated by observing the received signal after the last symbol duration and using that signal as part of the detection.

%

%\vspace{-0.3cm}
\subsection{Effects of Rapidly Changing Channels}
\label{subsec:changChan}

In this section we evaluate the performance of the SBRNN algorithm for rapidly changing channels. Due to lack of space we focus on the optical channel; we have observed similar performance results for the molecular channel as well. For modeling the rapidly changing channel, we assume that the channel parameter $\beta$ and the  noise parameter $\eta$ {\em change} from one symbol interval to the next. In particularly, we assume these parameters change according to a diffusion model with drift using the equations:
\begin{align}
	\beta_{i+1} &= \beta_{i} + d \beta_0 N    + \nu  \beta_0, \label{eq:betaDiffuse}\\
	\eta_{i+1} &= \eta_{i} + d \eta_0 N    + \nu  \eta_0,  \label{eq:etaDiffuse}
\end{align}
where $\beta_0$ and $\eta_0$ are the channel and noise parameters at the beginning of the transmission sequence, $d$ and $\nu$ control the diffusion and the drift velocities, and $N$ is a zero mean unit variance Gaussian random variable. The received signal is then given by
\begin{align}
\label{eq:diffusionPoisson}
\vec{y}_k[j] \sim \PoisDist\left( \sum_{i=0}^{k} x_{k-i} \bm{\lambda}_{i}^{\beta_{i}}[j] + \eta_k \right), 
\end{align}
where $\bm{\lambda}_{i}^{\beta_{i}}[j]$ is defined in \eqref{eq:impulseResOpti} and \eqref{eq:lambdaDefinition} with parameter $\beta_{i}$.

\begin{figure}
	\centering
	\includegraphics[width=0.9\columnwidth,keepaspectratio]{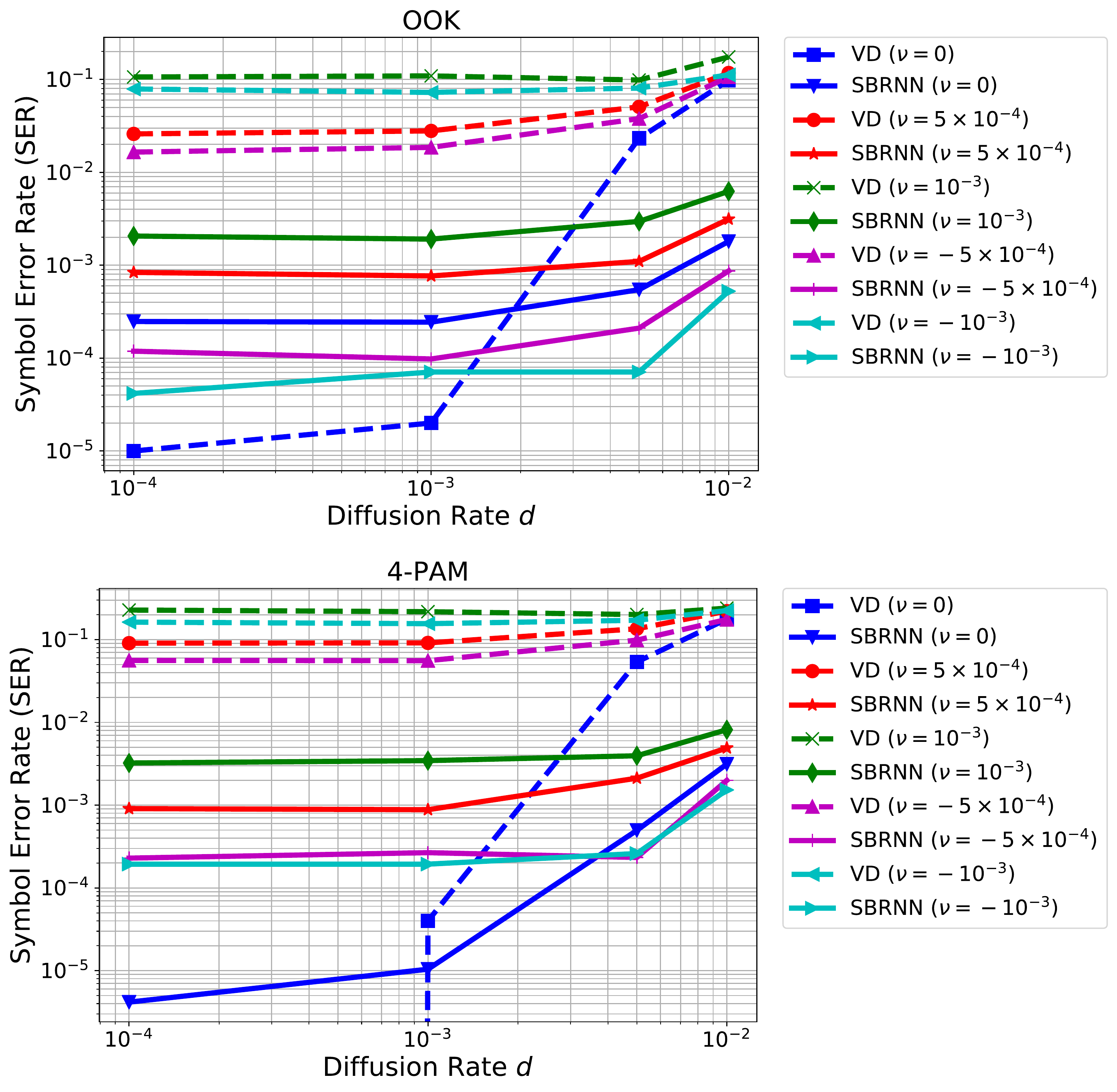}
	%\vspace{-0.3cm}
	\caption{\label{fig:diffuseResutls} The SBRNN performance under rapidly changing channel condition.}
	%\vspace{-0.9cm}
\end{figure}

The parameter $d$ controls the degree of dispersion, while the parameter $\nu$ controls how  $\beta$ and $\eta$ change on average over time. When $\nu=0$, $\mathbb{E}[\beta_i] = \beta_0$ and $\mathbb{E}[\eta_i] = \eta_0$. Note that $d>0$ controls the deviation from this mean. When $\nu>0$, the channel is degrading over time since $\mathbb{E}[\beta_i] > \beta_0$ and $\mathbb{E}[\eta_i] > \eta_0$, which result in larger ISI and noise components on average. Similarly, when $\nu<0$, the channel is improving over time because the ISI and the noise component are decreasing on average.

To evaluate the resiliency of the SBRNN detector to rapid changes in the channel, we use the same trained networks that were used to generate  Figures~\ref{fig:poisBERvSeqTimeEta}, \ref{fig:poisBERvBetaMuC}, and \ref{fig:4PAMTranLen}(a)-(b). Note that although these networks are trained using a data set that contains samples from various channel conditions, the channel parameters are fixed for the duration of the transmission of the whole sequence.  However, the model that is used for testing is the one in \eqref{eq:diffusionPoisson}, where the channel parameters changes from one symbol to the next during a transmission sequence. Specifically, for testing, sequences of length 200 symbols are used. The parameters of the channel are assumed to be $\beta_0=0.2$, $\eta_0=10$, and $\alpha=2$. For the OOK modulation $\tau=0.05$ $\mu$s, and $\kappa_{\text{OP}}=10$, while for 4-PAM $\tau=0.1$ $\mu$s, and $\kappa_{\text{OP}}=20$. The channel parameters $\beta_i$ and $\eta_i$ in \eqref{eq:diffusionPoisson} are assumed to diffuse according to \eqref{eq:betaDiffuse} and \eqref{eq:etaDiffuse} over a bounded intervals of $[0.15, 0.35]$ and $[1,200]$, respectively.

Figure \ref{fig:diffuseResutls} shows the results. For the VD plots, we assume that $\beta_0$ and $\eta_0$ is known perfectly at the receiver, i.e., the receiver has the perfect CSI at the beginning of the transmission sequence. If the diffusion rate is very small, and there is no drift (i.e., the channel is not changing), the VD performs very well, as expected. However, if the channel is drifting over time (i.e. $\nu>0$), the performance of the VD degrades significantly. Although the SBRNN algorithm is trained on a dataset where the channel does not change rapidly, it performs well under rapidly changing conditions. Also note that the training dataset has 100 symbol sequences while the test data has symbol sequences of length 200.  These results demonstrate that the SBRNN can be very useful in detection over rapidly changing channels, where traditional detection algorithms that cannot adapt well to the changing channel have performed poorly.

\vspace{-0.2cm}
\subsection{Computational Complexity}
\label{subsec:CompComplex}
We conclude this section by comparing the computational complexity of the SBRNN detector, the RNN detector, and the VD. Let $n$ be the length of the sequence to be decoded. Recall that $L$ is the length of the sliding BRNN, $M$ is the memory length of the channel, and $N$ is the number of states with the highest log-likelihood values among the $2^M$ states of the trellis that are kept at each time instance in the beam search Viterbi algorithm. Note that for the traditional Viterbi algorithm $N=2^M$. The computational complexity of the SBRNN is given by $O(L(n-L+1))$, while the computational complexity of the VD is given by $O(Nn)$. Therefore, for the traditional VD, the computational complexity grows exponentially with memory length $M$. However, this is not the case for the SBRNN detector. The computational complexity of the RNN detector is $O(n)$. Therefore, the RNN detector is the most efficient in terms of computational complexity, while the SBRNN detector and the beam search VD algorithm can have similar computational complexity. Finally, the traditional VD algorithm is impractical for the channels considered due to its exponential computational complexity in the memory length $M$.

\vspace{-0.2cm}
\section{Evaluation Based on Experimental Platform}
\label{sec:resultsExp}

In this section, we use a molecular communication platform for evaluating the performance of the proposed SBRNN detector. Note that although the proposed techniques can be used with any communication system, applying them to molecular communication systems enable many interesting applications. For example, one particular area of interest is in-body communication where bio-sensors, such as synthetic biological devices, constantly monitor the body for different bio-markers for diseases \cite{ata12CM}. Naturally, these biological sensors, which are adapt at detecting biomarkers {\em in vivo} \cite{and06,dan15,slo15}, need to convey their measurements to the outside world. Chemical signaling is a natural solution to this communication problem where the sensor nodes chemically send their measurements to each other or to other devices under/on the skin. The device on the skin is connected to the Internet through wireless technology and can therefore perform complex computations. Thus, the experimental platform we use in this work to validate NN algorithms for signal detection can be used directly to support this important application.   

We use the experimental platform in \cite{far17ExpPlat} to collect measurement data and create the dataset that is used for training and testing the detection algorithms. In the platform, time-slotted communication is employed where the transmitter modulates information on acid and base signals by injecting these chemicals into the channel during each symbol duration. The receiver then uses a pH probe for detection. A binary modulation scheme is used in the platform where the 0-bit is transmitted by pumping acid into the environment for 30 ms at the beginning of the symbol interval, and the 1-bit is represented by pumping base into the environment for 30 ms at the beginning of the symbol interval. The symbol interval consists of this 30 ms injection interval followed by a period of silence, which can also be considered as a guard band between symbols. In particular, four different silence durations (guard bands) of 220 ms, 304 ms, 350 ms, and 470 ms are used in this work to represent bit rates of 4, 3, 2.6, and 2 bps. This is similar to the OOK modulation used in the previous section for the Poisson channel model, except that chemicals of different types are released for both the 1-bit and the 0-bit.

To synchronize the transmitter and the receiver, every message sequence starts with one initial injection of acid into the environment for 100 ms followed by 900 ms of silence. The receiver then detects the starting point of this pulse by employing an edge detection algorithm and uses it to synchronize with the transmitter. Since the received signal is corrupted and noisy, this results in a random offset. However, since the NN detectors are trained directly on this data, as we will show, they learn to be resilient to this random offset. 

The training and test data sets are generated as follows. For each symbol duration, random bit sequences of length 120 are transmitted 100 times, where each of the 100 transmissions are separated in time. Since we assume no channel coding is used, the bits are i.i.d. and equiprobable.  This results in 12k bits per symbol duration that is used for training and testing. From the data, 84 transmissions per symbol duration (10,080 bits) are used for training and 16 transmissions are used for testing (1,920 bits). Therefore, the total number of training bits is 40,320, and the total number of bits used for testing is 7,680.

%We start by considering the slope detection using the rate of change of the pH. We use the training data to find the best detection parameters $B$ and $\gamma$, and the test data for evaluating the performance. In this section, we refer to this detection algorithm as the {\em baseline detection algorithm} since it was used in previous experimental implementations \cite{far13,koo16}. Besides this algorithm we consider different NN detectors. For all training, the Adam optimization algorithm \cite{kin14} is used with the learning rate $10^{-3}$.  Unless specified otherwise the number of epoch used during training is 200 and the batch size is 10. All the hyper parameters are tuned using grid search.
Although we expect from the physics of the chemical propagation and chemical reaction that the channel should have memory, since the channel model for this experimental platform is currently unknown, we implement both symbol-by-symbol and sequence detectors based on NNs. Note that due to the lack of a channel model, we cannot use the VD for comparison since it cannot be implemented without an underlying channel model. Instead, as a baseline detection algorithm, we use the slope detector that was used in previous work \cite{far13,koo16,far17ExpPlat}. For all training of the NN detectors, the Adam optimization algorithm \cite{kin14} is used with learning rate of $10^{-3}$.  Unless specified otherwise, the number of epochs used during training is 200 and the batch size is 10. All the hyperparameters are tuned using grid search.

We consider two symbol-by-symbol NN detectors. The first detector uses three fully connected layers with 80 hidden nodes and a final softmax layer for detection. Each fully connected layer uses the rectified linear unit (ReLU) activation function. The input to the network is a set of features extracted from the received signal, which are chosen based on performance and the characteristics of the physical channel as explained in the appendix. We refer to this network as {\em Base-Net}. A second symbol-by-symbol detector uses 1-dimensional CNNs. The best network architecture that we found has the following layers. 1) 16 filters of length 2 with ReLU activation; 2) 16 filters of length 4 with ReLU activation; 3) max pooling layer with pool size 2; 4) 16 filters of length 6 with ReLU activation; 5) 16 filters of length 8 with ReLU activation; 6) max pooling layer with pool size 2; 7) flatten and a softmax layer. The stride size for the filters is 1 in all layers. We refer to this network as {\em CNN-Net}.

For the sequence detection, we use three networks, two based on RNNs and one based on the SBRNN. The first network has 3 LSTM layers and a final softmax layer, where the length of the output of each LSTM layer is 40. Two different inputs are used with this network. In the first, the input is the same set of features as the Base-Net above. We refer to this network as {\em LSTM3-Net}. In the second, the input is the pretrained CNN-Net described above without the top softmax layer. In this network, the CNN-Net chooses the features directly from the received signal. We refer to this network as {\em CNN-LSTM3-Net}. Finally, we consider three layers of bidirectional LSTM cells, where each cell's output length is 40, and a final softmax layer. The input to this network is the same set of features used for Base-Net and the LSTM3-Net. When this network is used, during testing we use the SBRNN algorithm. We refer to this network as {\em SBLSTM3-Net}. For all the sequence detection algorithms, during testing, sample data sequences of the 120 bits are treated as an incoming data stream, and the detector estimates the bits one-by-one, simulating a real communication scenario. This demonstrates that these algorithms can work on any length data stream and can perform detection in real-time as data arrives at the receiver.

%For each of the networks, one parameter of interest that affects the input to the networks is the number of bins $B$. We have trained each network using different bin numbers to find the best value for each network. For the Base-Net $B=9$, for the CNN-Net $B=30$ and for all networks where the first layer is an LSTM or a BLSTM cell $B=8$. Note that during the training, for all NN detectors, the data from all symbol durations are used to train a single network, which can then perform detection on all symbol durations.

\begin{figure}
	\centering
	\includegraphics[width=0.75\columnwidth,keepaspectratio]{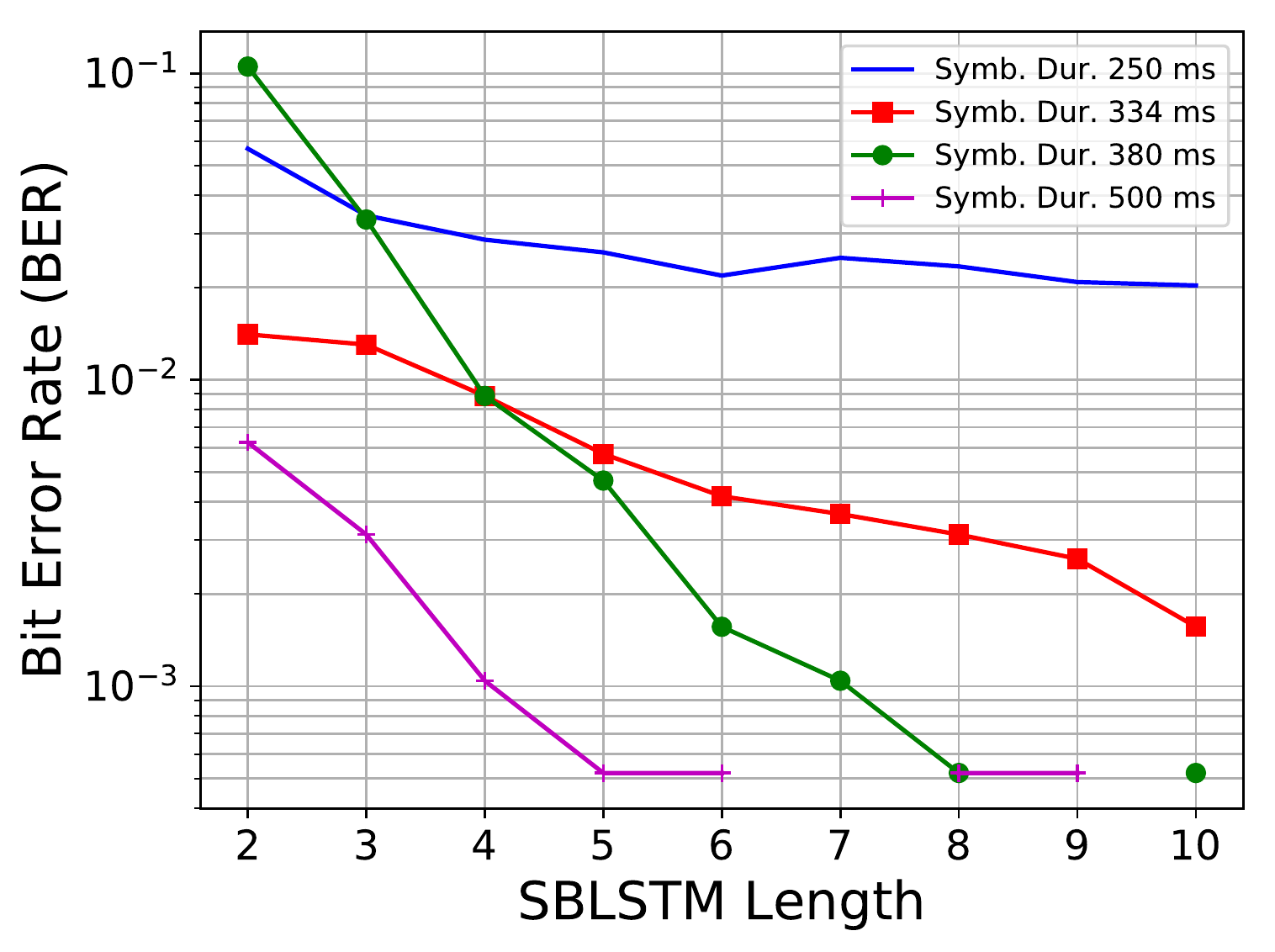}
	\vspace{-0.3cm}
	\caption{\label{fig:BERvSeqlen} The BER as a function of SBLSTM length.}
	\vspace{-0.4cm}
\end{figure}

\vspace{-0.3cm}
\subsection{System's Memory and ISI}
We first demonstrate that this communication system has a long memory. We use the RNN based detection techniques for this, and train the LSTM3-Net on sequences of 120 consecutive bits. The trained model is referred to as LSTM3-Net120. We run the trained model on the test data, once resetting the input state of the LSTM cell after each bit detection, and once passing the state as the input state for the next bit. Therefore, the former ignores the memory of the system and the ISI, while the latter considers the memory. The bit error rate (BER) performance for the memoryless LSTM3-Net120 detector is 0.1010 for 4 bps, and  0.0167 for 2 bps, while for the LSTM3-Net120 detector with memory, they are 0.0333 and 0.0005, respectively. This clearly demonstrates that the system has memory.

To evaluate the memory length, we train a length-10 SBLSTM3-Net on all sequences of 10 consecutive bits in the training data. Then, on the test data, we evaluate the BER performance for the SBLSTM of length 2 to 10. Figure~\ref{fig:BERvSeqlen} shows the results for each symbol duration. The BER reduces as the length of the SBLSTM increases, again confirming that the system has memory. For example, for the 500 ms symbol duration, from the plot, we conclude that the memory is longer than 4. Note that some of the missing points for the 500 ms and 380 ms symbol durations, which result in discontinuity in the plots, are because there were zero errors in the test data. Moreover, BER values below $5\times 10^{-3}$ are not very accurate since the number of errors in the test dataset are less than 10 (in a typical BER plot the number of errors should be about 100). However, given enough test data, it would be possible to estimate the channel memory using the SBLSTM detector by finding the minimum length after which BER does not improve.

% Baseline Technique Results on Good data
%[ 2.  5.  4.  3.]
%[ 0.  3.  1.  1.]
%[ 0.1296875   0.07552083  0.0796875   0.0515625 ]
% Baseline Technique Results on Bad data 20
%[ 4.  5.  5.  4.]
%[ 2.  3.  3.  1.]
%[ 0.15833333  0.13083333  0.125       0.07416667]

\vspace{-0.3cm}
\subsection{Performance and Resiliency}
\vspace{-0.1cm}
Table \ref{tb:BER} summarizes the best BER performance we obtain for all detection algorithms, including the baseline algorithm, by tuning all the hyperparameters using grid search. The number in front of the sequence detectors, indicates the sequence length. For example, LSTM3-Net120 is an LSTM3-Net that is trained on 120 bit sequences. In general, algorithms that use sequence detection perform significantly better than any symbol-by-symbol detection algorithm including the baseline algorithm. This is partly due to significant ISI present in the molecular communication platform. Overall, the proposed SBLSTM algorithm performs better than all other NN detectors considered. %Again note that BER values below $5\times 10^{-3}$ are not very accurate since the number of errors in the test dataset are less than 10, and more errors would be required for a better estimation of BER.

% SBLSTM 0.08833 0.0204166666667 0.05875 0.0141666666667

Another important issue for detection algorithms are changing channel conditions and resiliency. As the channel conditions worsen, the received signal is further degraded, which increases the BER. Although we assume no channel coding is used in this work, one way to mitigate this problem is by using stronger channel codes that can correct some of the errors. However, given that the NN detectors rely on training data to tune the detector parameters, overfitting may be an issue. To evaluate the susceptibility of NN detectors to this effect, we collect data with a pH probe that has a degraded response due to normal wear and tear. 

We collect 20 samples of 120 bit sequence transmissions for each of the 250 ms and 500 ms symbol durations using this degraded pH probe. First, to demonstrate that the response of the probe is indeed degraded, we evaluate it using the baseline slope-based detection algorithm. The best BERs obtained using the baseline detector are 0.1583 and 0.0741 for symbol durations of 250 ms and 500 ms, respectively. These values are significantly larger than those in Table~\ref{tb:BER}, because of the degraded pH probe. We then use the SBLSTM3-Net10 and the LSTM3-Net120, trained on the data from the good pH, on the test data from the degraded pH. For the SBLSTM3-Net10, the BERs obtained are 0.0883 and  0.0142, and for the LSTM3-Net120, the BERs are 0.1254 and 0.0504. These results confirm again that the proposed SBRNN algorithm is more resilient to changing channel conditions than the RNN. %Particularly, SBLSTM3-Net10's BER performance is several times better than the baseline, even though the baseline detector's parameters are refined based on the data collected using the degraded pH probe. This agrees with our results obtained using the Poisson channel model that the SBRNN is resilient to overfitting and changing channel conditions.   

\begin{table}[t]
	\scriptsize 
	\caption{Bit Error Rate Performance}
	\vspace{-0.3cm}
	\label{tb:BER}
	\centering
	\begin{tabular}{c|cccc}
		\toprule
		Symb. Dur. 			& 250 ms 		& 334 ms 		& 380 ms 		& 500 ms  \\
		\midrule
		Baseline			& 0.1297		& 0.0755  		& 0.0797 		& 0.0516  \\
		Base-Net   			& 0.1057		& 0.0245  		& 0.0380 		& 0.0115  \\
		CNN-Net    			& 0.1068    	& 0.0750  		& 0.0589 		& 0.0063  \\
		CNN-LSTM3-Net120	& 0.0677		& 0.0271		& 0.0026 		& 0.0021  \\
		LSTM3-Net120		&{\bf 0.0333}	& 0.0417		& 0.0083 		& 0.0005 \\
		SBLSTM3-Net10		& 0.0406		& {\bf 0.0141}	& {\bf 0.0005}	& {\bf 0.0000} \\
		\bottomrule
	\end{tabular}
	\vspace{-0.4cm}
\end{table}
\begin{table*}[t]
	\caption{Set of features that are extracted from the received signal and are used as input to different NN detectors in this paper. These values have been selected such that the trained network achieves the best result on a small validation set.}
	\label{tb:featureSets}
	\vspace{-0.2cm}
	\centering
	\begin{tabular}{l|ccccccc}
		\toprule
		Feature/Parameter 							& $B$	& $\gamma$  &$\vec{b}$	&$\vec{d}$	& $\hat{b}_0$ \& $\hat{b}_{B-1}$	& mean \& var $\vec{\hat{b}}$  &$\tau$\\
		\midrule
		Sec. \ref{sec:resultPoiss}: Optical Channel			& 10	& 1  		& No 		& Yes		& Yes								& Yes & Yes \\
		Sec. \ref{sec:resultPoiss}: Molecular Channel		& 10	& 1000 		& No 		& Yes		& Yes								& Yes & Yes \\
		Sec. \ref{sec:resultsExp}: Base-Net   		& 9		& 1 		& No 		& Yes		& Yes								& Yes & Yes  \\
		Sec. \ref{sec:resultsExp}: CNN-Net    		& 30	& 1 		& Yes 		& No		& No								& No & No\\
		Sec. \ref{sec:resultsExp}: CNN-LSTM3-Net120	& 30	& 1 		& Yes 		& No		& No								& No & No\\
		Sec. \ref{sec:resultsExp}: LSTM3-Net120		& 9		& 1 		& No 		& Yes		& Yes								& Yes & Yes\\
		Sec. \ref{sec:resultsExp}: SBLSTM3-Net10	& 9		& 1 		& No 		& Yes		& Yes								& Yes & Yes\\
		\bottomrule
	\end{tabular}
	\vspace{-0.5cm}
\end{table*}

Finally, to demonstrate that the proposed SBRNN algorithm can be implemented as part of a real-time communication system, we use it to support a text messaging application built on top of the experimental platform. We demonstrate that using the SBRNN for detection at the receiver, we are able to reliably transmit and receive messages at 2 bps. This data rate is an order of magnitude higher than previous systems \cite{far13,koo16}.

\section{Conclusions}
\label{sec:conclusion}
This work considered a machine learning approach to the detection problem in communication systems. In this scheme, a neural network detector is directly trained using measurement data from experiments, data collected in the field, or data generated from channel models. Different NN architectures were considered for symbol-by-symbol and sequence detection. For channels with memory, which rely on sequence detection, the SBRNN detection algorithm was presented for real-time symbol detection in data streams. To evaluate the performance of the proposed algorithm, the Poisson channel model for molecular communication was considered as well as the VD for this channel. It was shown that the proposed SBRNN algorithm can achieve a performance close to the VD with perfect CSI, and better than the RNN detector and the VD with CSI estimation error. Moreover, it was demonstrated that using a rich training dataset that contains sample transmission data under various channel conditions, the SBRNN detector can be trained to be resilient to the changes in the channel, and achieves a good BER performance for a wide range of channel conditions. Finally, to demonstrate that this algorithm can be implemented in practice, a molecular communication platform that uses multiple chemicals for signaling was used. Although the underlying channel model for this platform is unknown, it was demonstrated that NN detectors can be trained directly from experimental data. The SBRNN algorithm was shown to achieve the best BER performance among all other considered algorithms based on NNs as well as a slope detector considered in previous work. Finally, a text messaging application was implemented on the experimental platform for demonstration where it was shown that reliable communication at rates of 2 bps is possible, which is an order of magnitude faster than the data rate reported in previous work for molecular communication channels.

As part of future work we plan to investigate how techniques from reinforcement learning could be used to better respond to changing channel conditions. We would also like to study if the evolution of the internal state of the SBRNN detector could help in developing channel models for systems where the underlying models are unknown.

\appendices

%%%%%%%%%%%%%%%%%%%%%%%%%%%%%%%%%%%%%%%%%%%%%%%%%%%%%%%%%%%%%%%%%%%%%%%%%%%%%%%%%%%%%%
\vspace{-0.3cm}
\section*{Appendix\\Feature Extraction} %\label{annex:featureExt}
In this appendix we describe the set of features that are extracted from the received signal and are used as the input to the different NN detectors considered in this work. The set of features $\vec{r}_k$, extracted from the received signal during the~\kth channel use $\vec{y}_k$, must preserve and summarize the important information-bearing components of the received signal. For the Poisson channel, since the information is encoded in the intensity of the signal, much of the information is contained in the rate of change of intensity. In particular, intensity increases in response to the transmission of the 1-bit, while intensity decreases or remains the same in response to transmission of the 0-bit. Note that this is also true for the pH signal in the experimental platform used in Section \ref{sec:resultsExp}. First the symbol interval (i.e., the time between the green lines in Figure~\ref{fig:rcvSig}) is divided into a number of equal subintervals or bins. Then the values inside each bin are averaged to represent the value for the corresponding bin. Let $B$ be the number of bins, and $\vec{b}=[b_0,b_1,\cdots,b_{B-1}]$ the corresponding values of each bin. We then extract the rate of change during a symbol duration by differentiating the bin vector to obtain the vector $\vec{d}=[d_0,d_1,\cdots,d_{B-2}]$, where $d_{i-1} = b_i- b_{i-1}$. We refer to this vector as the {\em slope vector} and use it as part of the feature set $\vec{r}_k$ extracted from the received signal. 

Other values that can be used to infer the rate of change are $b_0$ and $b_{B-1}$, the value of the first and the last bins, and the mean and the variance of the $\vec{b}$. Since the intensity can grow large due to ISI, $\vec{b}$ may be normalized with the parameter $\gamma$ as $\vec{\hat{b}}=\vec{b}/\gamma$. Therefore, instead of $b_0$ and $b_{B-1}$, $\hat{b}_0$ and $\hat{b}_{B-1}$, and the mean and the variance of the $\vec{\hat{b}}$ may be used as part of the feature set $\vec{r}_k$. Finally, since the transmitter and the receiver have to agree on the symbol duration, the receiver knows the symbol duration, which can be part of the feature set.  Table \ref{tb:featureSets} summarizes the set of features that are used as input to the each of the NN detection algorithms in this paper.

%\appendices
%% you can choose not to have a title for an appendix
%% if you want by leaving the argument blank

%% use section* for acknowledgment
%\ifCLASSOPTIONcompsoc
%  % The Computer Society usually uses the plural form
%\else
%  % regular IEEE prefers the singular form
%\fi

\ifCLASSOPTIONcaptionsoff
  \newpage
\fi



%\renewcommand{\bibfont}{\tiny}
%\vspace{-0.4cm}
\bibliographystyle{IEEEtran}
\bibliography{IEEEabrv,MolCom-Nariman,ML-Nariman}
\end{document}